\documentclass[journal]{IEEEtran}
\usepackage{amsmath} 
\usepackage{amssymb}
\usepackage{amsfonts}
\usepackage{amsthm}
\allowdisplaybreaks[4]
\usepackage{algorithmic}
\usepackage{algorithm}
\usepackage{gensymb}
\usepackage{array}
\usepackage{booktabs}
\usepackage[caption=false,font=normalsize,labelfont=sf,textfont=sf]{subfig}
\usepackage{textcomp}
\usepackage{color}
\usepackage{bm}
\usepackage{stfloats}
\usepackage{caption}
\usepackage{url}
\usepackage{verbatim}
\usepackage{graphicx}
\usepackage{diagbox}
\usepackage{epstopdf}
\usepackage{cite}

\newtheorem{lemma}{Lemma}

\newtheorem{remark}{Remark}

\begin{document}

\title{Enhancing Robustness and Security in ISAC Network Design: Leveraging Transmissive Reconfigurable Intelligent Surface with RSMA}

\author{Ziwei~Liu,~Wen~Chen,~Qingqing~Wu,~Zhendong~Li,~Xusheng~Zhu,~Qiong~Wu,~and~Nan Cheng
	

	\thanks{Z. Liu, W. Chen, Q. Wu, and X. Zhu are with the Department of Electronic Engineering, Shanghai Jiao Tong University, Shanghai 200240, China (e-mail: ziweiliu@sjtu.edu.cn; wenchen@sjtu.edu.cn; qingqingwu@sjtu.edu.cn;  xushengzhu@sjtu.edu.cn).}
	\thanks{Z. Li is with the School of Information and Communication Engineering, Xi'an Jiaotong University, Xi'an 710049, China (e-mail: lizhendong@xjtu.edu.cn).}
	\thanks{Q. Wu is with the School of Internet of Things Engineering, Jiangnan
	University, Wuxi 214122, China (e-mail: qiongwu@jiangnan.edu.cn). }
	\thanks{N. Cheng is with the State Key Lab. of ISN and School of Telecommunications Engineering, Xidian University, Xi'an 710071, China (e-mail: dr.nan.cheng@ieee.org).	}
		

}

%

\maketitle

\begin{abstract}
In this paper, we propose a novel transmissive reconfigurable intelligent surface (TRIS) transceiver-enhanced robust and secure integrated sensing and communication (ISAC) network. A time-division sensing communication mechanism is designed for the scenario, which enables communication and sensing to share wireless resources. To address the interference management problem and hinder eavesdropping, we implement  rate-splitting multiple access (RSMA), where the common stream is designed as a useful signal and an artificial noise (AN), while taking into account the imperfect channel state information and modeling the channel for the illegal users in a fine-grained manner as well as giving an upper bound on the error. We introduce the secrecy outage probability and construct an optimization problem with secrecy sum-rate as the objective functions to optimize the common stream beamforming matrix, the private stream beamforming matrix and the timeslot duration variable. Due to the coupling of the optimization variables and the infinity of the error set, the proposed problem is a nonconvex optimization problem that cannot be solved directly. In order to address the above challenges, the block coordinate descent (BCD)-based second-order cone programming (SOCP) algorithm is used to decouple the optimization variables and solving the problem. Specifically, the problem is decoupled into two subproblems concerning the common stream beamforming matrix, the private stream beamforming matrix, and the timeslot duration variable, which are solved by alternating optimization until convergence is reached. To solve the problem, S-procedure, Bernstein's inequality and successive convex approximation (SCA) are employed to deal with the objective function and non-convex constraints. Numerical simulation results verify the superiority of the proposed scheme in improving the secrecy energy efficiency (SEE) and the Cram\'{e}r-Rao boundary (CRB).
\end{abstract}

\section{Introduction}
\IEEEPARstart{I}{n} the realm of modern communications, ensuring the security and confidentiality of transmitted data is a paramount concern.  While traditional cryptographic techniques have historically focused on securing data at higher protocol layers, in recent years both academia and industry have become increasingly interested in strengthening the foundation of communication systems, the physical layer, due to its ability to protect data confidentiality without relying on key distribution or encryption/decryption, as well as latency advantages over bit-level cryptographic techniques\cite{e19080420,9702524,10736549}. However, the inherent trade-offs between security, performance and compatibility, as well as the issues of signal leakage and interference management, constrain the practical deployment of physical layer security communication (PLSC) solutions in various communication scenarios\cite{9762838}. 

The reconfigurable intelligent surface (RIS), a new paradigm for enhancing the security of the physical layer of wireless networks, significantly impacts the landscape of wireless communication systems, providing opportunities to enhance data transmission and against threats\cite{10188924,9941040,10463696,wu2021intelligent}. One of the main mechanisms by which RIS contributes to physical layer security is channel conditioning, i.e., by controlling the phase shift of the reflected signals, RIS can reshape the wireless propagation environment to create favorable conditions for the desired communication links while reducing eavesdropping attempts. Z. Li {\it et. al} investigated the problem of deploying RIS to reconfigure the wireless channel and maximizing the secrecy energy efficiency under outage probability constraints and eavesdropper equipped with multi-antenna conditions, and results demonstrate the importance of optimizing the RIS phase shift to defend against eavesdropping\cite{10320405}. Subsidiarily, RIS facilitates the implementation of secure beamforming techniques that can selectively increase the signal-to-noise ratio (SNR) for legitimate users while limiting signal leakage and blocking unauthorized reception. Y. Sun {\it et. al} investigated the role of RIS in improving the spectral efficiency and security of multiuser cellular networks by converting imperfect angle channel state information (CSI) to robust CSI and designing robust RIS hybrid beamforming to solve the worst-case and rate maximization problems\cite{9779086}. In addition, RIS relies on its tunable elements to enable functions such as signal enhancement and optimization of signal paths in specific directions, which can be used for assisted sensing to achieve precise positioning of targets\cite{10507060,10056405,10437949}. Recently, a new type of transmissive RIS (TRIS) transceiver has entered the research horizon\cite{10680462}. This transceiver can realize the function of multi-antenna system in the form of low energy consumption, and it is a promising wireless solution with higher aperture efficiency, finer beamforming and avoiding the problem of reflected wave interference compared with the reflective RIS\cite{10740042,10242373}. Due to its special structural characteristics, how its power constraints and received signal processing can be adapted to a specific network is a problem yet to be solved. However, RIS is highly dependent on accurate CSI and there is a growing need for proactive sensing of potential threats in wireless environments, so further improvements in this area are essential\cite{10143420}.

ISAC technology is centered on playing a key role in providing communication and understanding the background of the wireless environment, thus providing a solution for security mechanisms. By collecting CSI, monitoring signal strength and anomaly detection, security mechanisms can proactively identify and respond to security threats, thereby protecting communication integrity and preserving network confidentiality\cite{Su2023,9755276,9237455,10464353,zhang2024multipleintelligentreflectingsurfaces}. Z. Ren {\it et. al} investigated the secure communication sensing problem in both cases of bounded CSI errors and CSI errors conforming to a Gaussian distribution, where the base station (BS) transmits a confidential signal integrating sensing signals to the communication user, while sensing targets that may be suspected eavesdroppers\cite{10153696}. In order to explore in depth the impact of channel errors, an eavesdropping channel error safe approximation problem is investigated in \cite{9933849}, and secure communication and sensing is achieved by utilizing the variable-length snapshot framework. Sense-assisted acquisition of eavesdropper information is valuable for secure communications. The BS first transmits omnidirectional waveforms to obtain potential Eves information, based on which a secrecy rate expression is formulated, and then, constructs the optimization problem for simultaneously maximizing the secrecy rate and minimizing the target/Eves estimation of the Cram\'{e}r-Rao boundary (CRB)\cite{10227884}. By improving the estimation accuracy, the sensing and security functions are mutually beneficial. H. Jia {\it et. al} considered dual-functional radar communications with spectral and energy-efficient characteristics to enhance PLSC under eavesdropper CSI uncertainty, and utilizes the CRB as a PLSC-optimized model for the sensing metric while satisfying the tolerance sensing requirement and the secrecy rate constraint\cite{10375133}. However, The problem of interference management is a significant challenge due to malicious attacks by eavesdroppers and the mixing of sensing and communication signals, and solving this challenge is the key to unleashing the full potential of ISAC\cite{10663785}.

Rate splitting multiple access (RSMA) is known to better manage interference and thus achieve higher spectral efficiency\cite{9831440,9832611,Mao2018}. By shaping the transmitted waveforms and processing the received signals, RSMA can suppress interference from other signal sources and improve the SNR, thus enhancing the robustness and reliability of the radar system. In addition, the diversification of transmission signals by splitting transmission data into common and private streams makes it more difficult for potential adversaries to intercept or decipher sensitive information, while RSMA networks share common information stream and co-optimize the allocation of resources, which can be used for identifying abnormal behaviors and threats utilizing collective intelligence. In \cite{xia2023weighted}, an RSMA-based secure beamforming method is proposed to maximize the weighted sum-rate. Simulations show performance superiority in channel error robustness and interference management compared to the baseline. In \cite{9967957}, the confidentiality performance of RSMA in a multiuser multiple-input single-output system is investigated, and simulations show that by adjusting the partitioning of the messages, the proposed power allocation methodology allows for a scalable trade-off between rate effectiveness and confidentiality. Furthermore, the common stream of RSMA is used for sensing, based on which the authors construct and maximize a performance criterion for sensing, with guaranteed communication\cite{10522473}. In conventional ISAC systems, multiplicative fading of radar signals and space division multiple access limit the system performance. A novel RSMA-RIS-assisted ISAC system is proposed in \cite{10287099} to improve sensing performance. RSMA has more flexible interference management capabilities due to its additional common stream, which provides potential design freedom. 

Overall, the user's demand for data security imposes higher requirements on future network design, which is accompanied by interference management and energy consumption issues. In this paper, we employ TRIS to empower ISAC networks and provide a low-cost, low-energy architecture in which RSMA is employed to manage interference among users. To the best of our knowledge, there is few research on TRIS-enabled ISAC networks. According to the scenario requirements, an optimization problem is constructed under imperfect CSI conditions with maximizing secrecy sum-rate as the objective function, while setting secrecy outage as a constraint to further improve the secrecy of the network. The main contributions of this paper are as follows:

\begin{itemize}
\item[$\bullet$] We propose a time-division secure ISAC architecture. In this architecture, communication and sensing share the same wireless resources, and the wireless environment information obtained by sensing brings benefits to communication while detecting potential illegal users (IUs). To facilitate the integration of sensing and communication, we deploy a novel transmissive reconfigurable intelligent surface (TRIS) transceiver framework that utilizes time modulatin array (TMA) for simultaneous multistream communication and sensing, and innovatively exploits the common stream for RSMA as both useful signals and artificial noise, which has not been considered in previous work.
\item[$\bullet$] Based on the architectural setting, we consider the problem of secure communication and detection of potentially IUs under conditions of imperfect CSI and network serving multiple legitimate users (LUs) with the presence of multiple IUs. Since accurate estimation of IU channel is quite challenging, we model the channel and estimation error of IUs in a refined way and give theoretical upper bounds on the error. Concurrently, to improve the security of the system, we consider system outage when the secrecy rate falls below a threshold.
\item[$\bullet$] In response to the above problems, we construct an optimization problem taking secrecy sum-rate as the objective function and beampattern, detection probability and outage probability as the constraints. The nonconvexity of the objective function and constraints are handled by using S-procedure, Bernstein's inequality and successive convex approximation (SCA). Finally, the optimization problem is decoupled and solved using block coordinate descent (BCD) algorithm and second order convex cone programing (SOCP) algorithm. We evaluate the performance and numerical simulations show that the proposed scheme has a minimum gain in secrecy energy efficiency (SEE) of 44\% compared to traditional transceiver, confirming the superiority of the proposed scheme.
\end{itemize}

\emph{Notations}: Scalars are denoted by lower-case letters, while vectors and matrices are represented by bold lower-case letters and bold upper-case letters, respectively. $|x|$ denotes the absolute value of a complex-valued scalar $x$, ${x^ * }$ denotes the conjugate operation, and $\left\| \bf x \right\|$ denotes the Euclidean norm of a complex-valued vector $\bf x$. For a square matrix ${\bf{X}}$, ${\rm{tr}}\left( {\bf{X}} \right)$, ${\rm{rank}}\left( {\bf{X}} \right)$, ${{\bf{X}}^H}$, ${\left[ {\bf{X}} \right]_{m,n}}$ and $\left\| {\bf{X}} \right\|$ denote its trace, rank, conjugate transpose, ${m,n}$-th entry, and matrix norm, respectively. ${\bf{X}} \succeq 0$ represents that ${\bf{X}}$ is a positive semidefinite matrix. In addition, ${\mathbb{C}^{M \times N}}$ denotes the space of ${M \times N}$ complex matrices. $j$ denotes the imaginary element, i.e., $j^2 = -1$. The distribution of a circularly symmetric complex Gaussian (CSCG) random vector with mean $\mu $ and variance $\sigma^2$ is denoted by ${\cal C}{\cal N}\left( {{\mu},\sigma^2} \right)$ and $ \sim $ stands for ‘distributed as’. ${\bf{A}} \otimes {\bf{B}}$ represents the Kronecker product of matrices ${\bf{A}}$ and ${\bf{B}}$.
\begin{figure}[H]
		\centerline{\includegraphics[width=9cm]{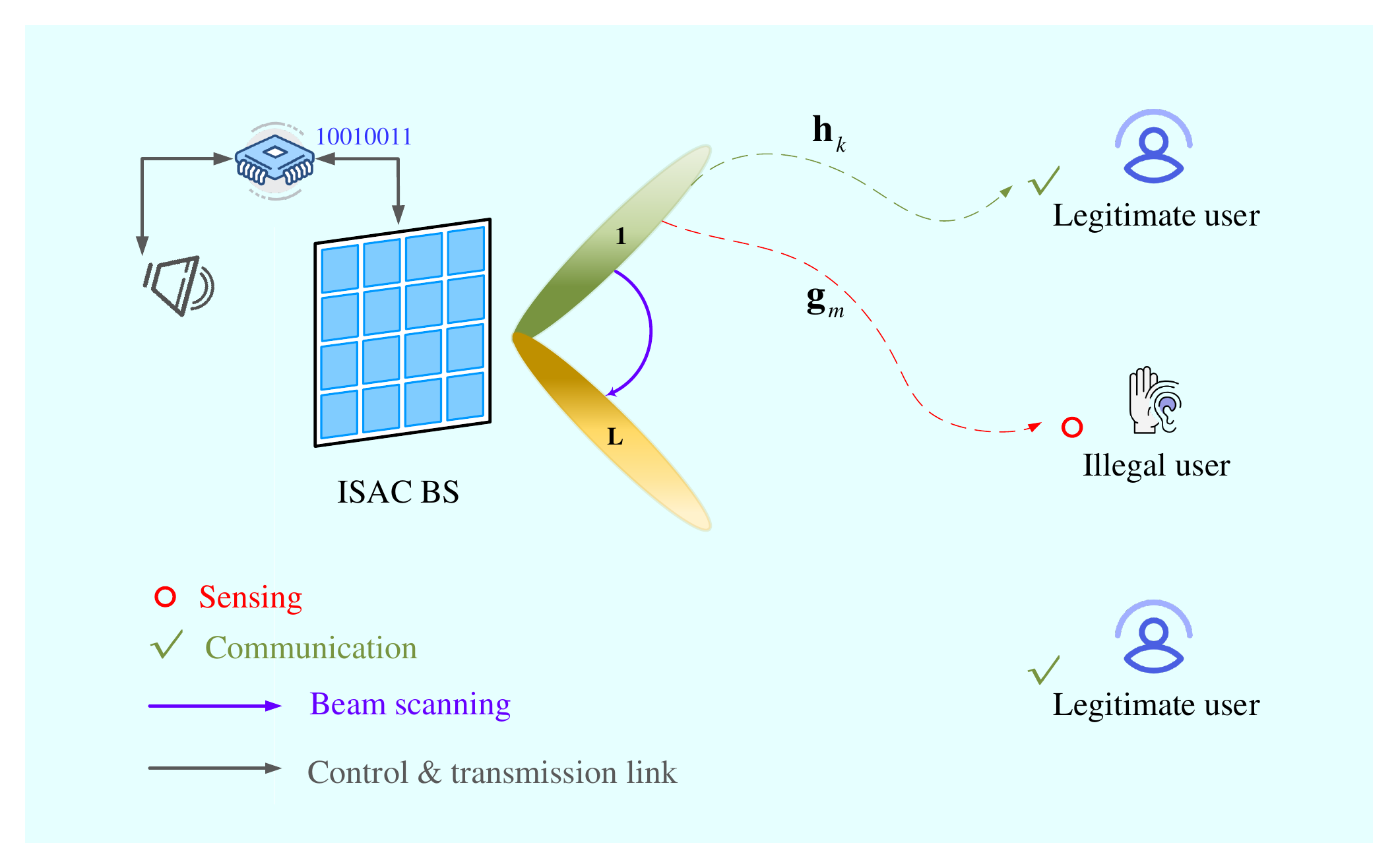}}
	\caption{TRIS transceiver empowered ISAC networks.}\label{sys}
\end{figure}
\section{System Model}
In this section, we first present a model of the ISAC network employing the TRIS transmitter and give the sensing mechanism. Next, based on the characteristics of this transmitter, the corresponding channel models are established and the imperfect knowledge of CSI of LUs and IUs is taken into account. Finally, the corresponding signal models are given.
\subsection{ISAC Network Models}
We consider a downlink multiuser ISAC system consisting of a TRIS-empowered BS with $K$ LUs and $M$ IUs\footnote{The information about the LUs is known in advance by the BS but is not accurate and the information about the IUs such as their location is to be obtained. The availability of this information allows for better beam alignment and thus improved security, the specific process is given in the secure mechanism, as detailed in Fig. 3 and the description below.}, as shown in Fig. \ref{sys}. The BS consists of a TRIS with $N = {N_r} \times {N_c}$ elements arranged in a uniform planar array (UPA) pattern, a horn antenna and a controller\footnote{In this structure, since the antenna and the users are located on different sides of the TRIS, the feed source obstruction problem and the self-interference of the reflected wave in the reflected RIS are solved. Furthermore, TRIS serves as a spatial diversity and loading information, which utilizes the control signals generated by the TMA to load precoded information onto the TRIS elements, and each element is loaded with different information, and the carrier wave penetrates the elements and carries the information to achieve direct modulation, thus realizing the function of traditional multiple antennas with lower energy consumption\cite{Liu2023,10740042}. Besides, TMA provides a solution for multi stream communication for this architecture.}, while LUs and previously detected IUs are single antenna devices. 
\begin{figure}[H]
	\centerline{\includegraphics[width=9cm]{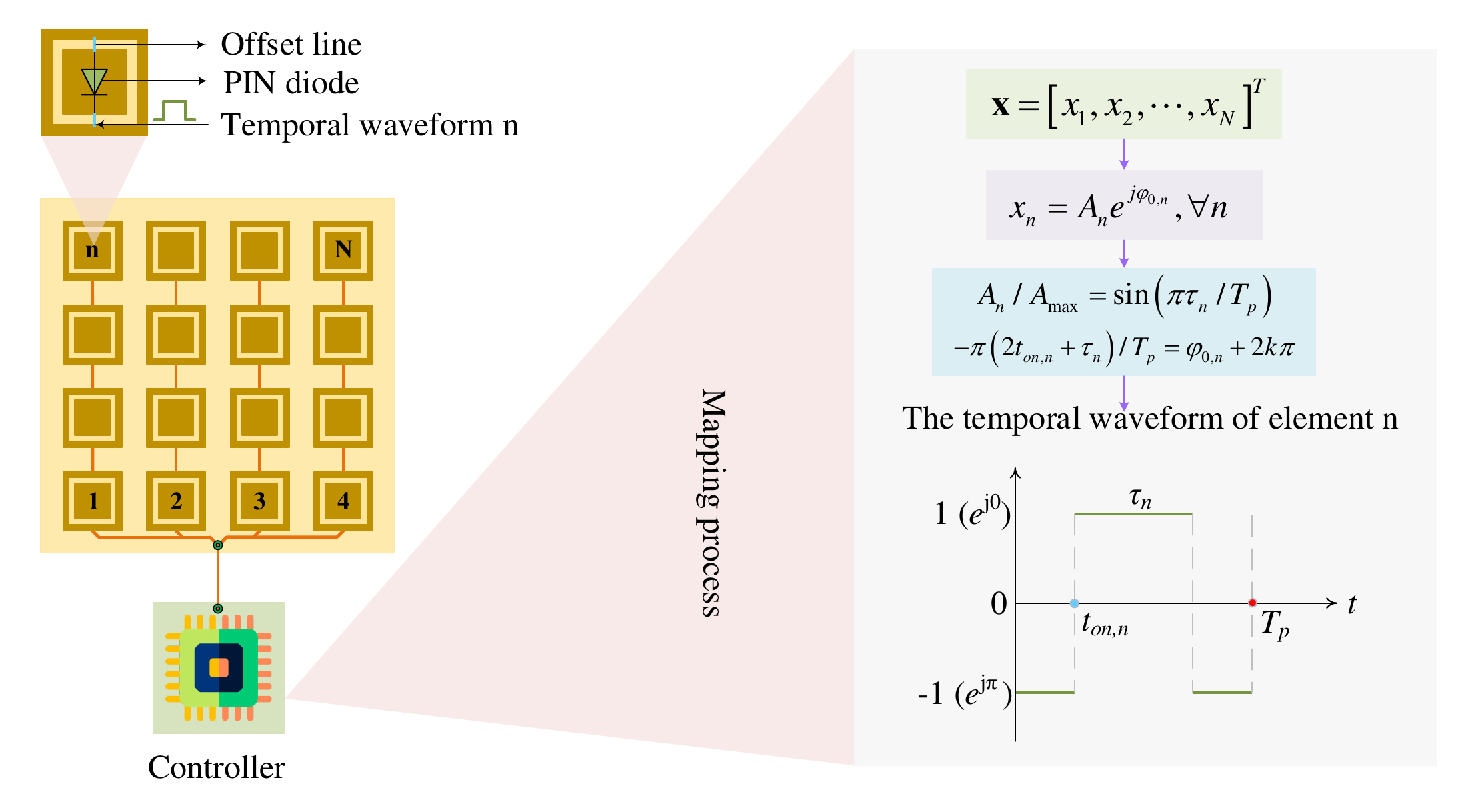}}
	\caption{The principle and mapping process of TRIS.}\label{pri}
\end{figure}
In this paper, the TRIS transceiver is utilized because it can realize the function of traditional multiple antennas in the form of low power consumption, and it can realize finer beamforming by using TMA, and it has a larger aperture efficiency than reflective RIS and avoids the interference problem of reflected waves, so it has a key role in improving the performance of secure communication and sensing accuracy and also is more suitable for the future network of the low-power demand. Since this paper uses a new type of transmitter, the precoding matrix needs to meet the design requirements of the transmitter. Based on literature \cite{he}, the transmit signal is modulated on positive and negative 1st harmonics, there will be a power loss of 0.91 dB, and the power of the signal loaded on the TRIS element cannot exceed the available power of TRIS element, so the power constraints need to satisfy the following expression\cite{10740042}.

\begin{equation}
			\setlength{\abovedisplayskip}{3pt}
	\setlength{\belowdisplayskip}{3pt}
{{{\left[ {{{\bf{W}}\left[ l \right]}{\bf{W}}^H\left[ l \right]} \right]}_{nn}}}  \le {P_t},\forall n,
\end{equation}
where ${\bf{W}}\left[ l \right]$ denotes the beam matrix in time slot $l$, and the maximum available power of each TRIS element is $P_t$. The reason that the power constraint is written as Eq. (1) is that in the TMA modulation process, the complex value of each element $x_n=A_ne^{j\varphi_n}$ of the signal ${\bf x}\left[l\right]={\bf W}\left[l\right]{\bf s}\left[l\right]\in \mathbb{C}^{N \times l}$ is first modulated to the 1st harmonic according to Eqs. (2) and (3), then transformed to a temporal waveform, and finally loaded onto the TRIS elements. Therefore it is required that the power of element $x_n$ should not exceed the available power of the TRIS element. Based on the matrix multiplication rules, this requires constraints on the rows of the precoding matrix. 

\begin{equation}
			\setlength{\abovedisplayskip}{3pt}
	\setlength{\belowdisplayskip}{3pt}
		A_n/{A_{\max }} = \sin \left( {\pi \tau_n /{T_p}} \right),
\end{equation}
and
\begin{equation}
			\setlength{\abovedisplayskip}{3pt}
	\setlength{\belowdisplayskip}{3pt}
	- \pi \left( {2{t_{on,n}} + \tau_n } \right)/{T_p} = {\varphi _{0,n}} + 2k\pi ,\forall k,
\end{equation} 
where ${T_p}$ is the code element time, ${A_{\max }}$ is the maximum amplitude of the modulated signal, $t_{on,n}$ denotes the moment of 0-state onset, and $\tau_n$ indicates the duration of the 0 state. The principle and mapping process are shown in Fig. \ref{pri}. 

\begin{figure*}
	{\centerline{\includegraphics[width=18cm]{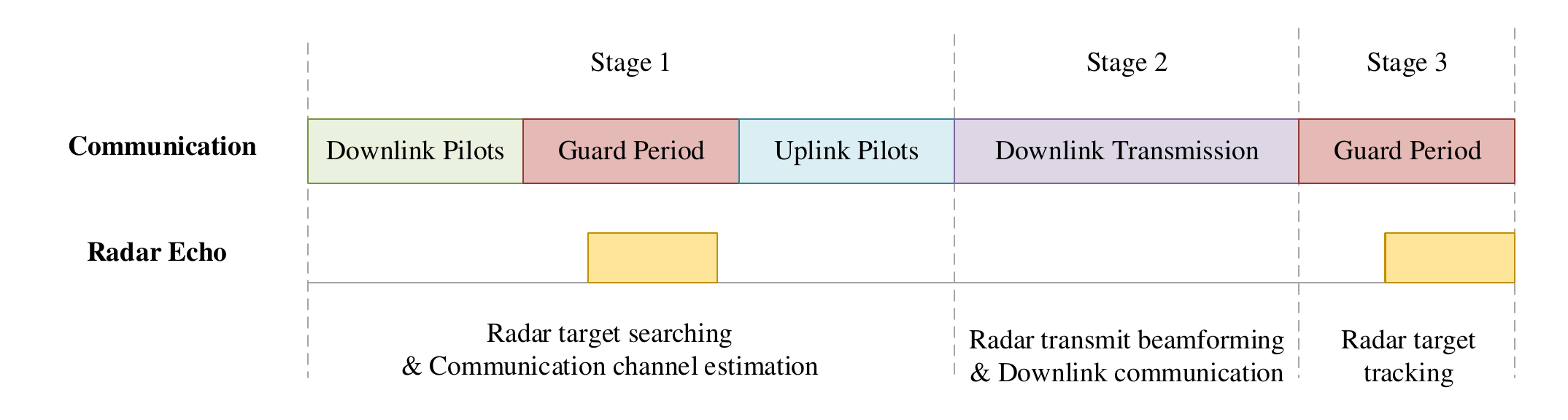}}}
	\caption{Frame structure of the ISAC network.}\label{arc}
\end{figure*}

Regarding the secure ISAC mechanism, its frame structure can be divided into three stages as shown in Fig. \ref{arc}. {\bf (1) Radar target searching and communication channel estimation}: In this stage, the BS first transmits an omnidirectional waveform containing the downlink pilots, and then estimates the parameters such as angle of arrival, distance, and Doppler in the echo while detecting all targets. Next, the LUs receive the probing signal and estimates the departure of angle and send the uplink pilots to the BS. Meanwhile, the IUs passively transmit their geometrical parameters to the BS by reflecting probing signals\footnote{In this paper, the access method utilizes RSMA, where the BS transmits an integrated waveform for communication and sensing, and in addition the common stream is used as an AN for the IUs. the common stream as an AN is justified because the LUs are authorized to interact with the BS via the uplink pilot and all LUs share both the encoding and decoding rules for the common stream so that the IUs are not able to decode the information. In addition, both LUs and IUs are considered as targets, and the BS distinguishes between them by actively or passively interacting with the two.}. Finally, the BS can obtain the channel parameters and detect active LUs as well as potential IUs. {\bf (2) Radar transmit beamforming and downlink communication}: In this stage, the BS has acquired the CSI containing the angle information and sends the beam in the direction of interest through the co-design of the communication-sensing beamformer for further detection of potentially IUs and downlink communication. {\bf (3) Radar target tracking}: In this stage, the BS receives the echoes from the previous stage to update the target parameters. In this paper, we focus on stage 2, and investigates wireless resource allocation and optimization in this stage, where the BS transmits a total of ${\cal L}=\left\{1,\cdots,L\right\}$ beam sequences at different frames and scans a sector during scanning period $T$ to downlink communication while detecting potential new IUs.

\subsection{Channel Model} 

\subsubsection{Communication Channels for LUs} The communication channel ${{\bf{h}}_k\left[ l \right]} \in {\mathbb{C}^{N \times 1}}$ between the LU $k$ and the BS consists of a line-of-sight (LoS) portion and a non-line-of-sight (NLoS) portion, which can be modeled as the following Rician fading channel
\begin{equation}
{{\bf{h}}_k}\left[ l \right] = {\xi _k}\left[ l \right]\left( {\sqrt {\frac{{{\kappa _v}}}{{{\kappa _v} + 1}}} {\overline {\bf{h}} _k}\left[ l \right] + \sqrt {\frac{1}{{{\kappa _v} + 1}}} {\underline {\bf{h}}_k}\left[ l \right]} \right),\forall k,\label{h}
\end{equation}
where ${\xi _k\left[ l \right]={{\lambda _c}}/{{4\pi {d_k}\left[ l \right]}}}$ denotes the path loss and ${\kappa _v}$ represents the Rician factor. ${\lambda _c}$ represents the carrier wavelength, and ${d_k}\left[ l \right]$ represents the distance between the LU $k$ and the BS. The LoS channel can be expressed as

\begin{equation}
{\overline {\bf{h}} _k}\left[ l \right] = {\left[ {{e^{ - j2\pi {\delta _r}{{\bf{n}}_r}}}} \right]^T} \otimes {\left[ {{e^{ - j2\pi {\delta _c}{{\bf{n}}_c}}}} \right]^T},\forall k,
\end{equation}
where ${{\bf{n}}_r} = \left[ {0,1, \cdots ,{N_r} - 1} \right]$, ${{\bf{n}}_c} = \left[ {0,1, \cdots ,{N_c} - 1} \right]$, ${\delta _r} = {{d_f}\sin {\theta _k}\left[ l \right]\cos {\varphi _k}\left[ l \right]}/{\lambda_c }$, and ${\delta _c} = {{d_f}\sin {\theta _k}\left[ l \right]\sin {\varphi _k}\left[ l \right]}/{\lambda_c }$. $\left( {{n_r},{n_c}} \right)$ denotes the element position index of the RIS and ${d_f}$ denotes the center distance of adjacent RIS elements.  ${\varphi _k}\left[ l \right]$ and ${\theta _k}\left[ l \right]$ denote the user's azimuth and pitch angles, respectively. The NLoS channel obeys a CSCG distribution, i.e., ${\underline {\bf{h}}_k}\left[ l \right] \sim {\cal C}{\cal N}\left( {0,{{\bf{I}}_N}} \right)$. 

In addition, there is often uncertainty in the channel due to user movement, scattering from the surrounding environment, and scanning period, and in order to portray this uncertainty, an uncertainty model is used to quantify the error in the channel. The channel error vector for the LU $k$ can be expressed as

\begin{equation}
			\setlength{\abovedisplayskip}{3pt}
	\setlength{\belowdisplayskip}{3pt}
\Delta {{\bf{h}}_k}\left[ l \right] = {{\bf{h}}_k}\left[ l \right] - {\widehat {\bf{h}}_k},\forall k,
\end{equation}
where ${\widehat {\bf{h}}_k}$ denotes the estimated channel of LU $k$ before the scanning period and ${{\bf{\hat h}}_k}$ and $\Delta {{\bf{h}}_k}\left[ l \right]\sim {\cal C}{\cal N}\left( {0,{{\sigma_{h_e}^2}{\bf{I}}_N}} \right)$ are independent. ${\widehat {\bf{h}}_k}$ can be characterized by Eq. (\ref{h}) and is assumed to be known by the BS, and can be obtained by the CSI algorithm\cite{9839429}. For LUs, who interact with the BS before beam scanning, more accurate CSI can be obtained therefore $\Delta {{\bf{h}}_k}\left[ l \right]$ is usually bounded, which can be constrained by the bounded uncertainty model, i.e., $\left\| \Delta {{\bf{h}}_k}\left[ l \right] \right\| \le {\varepsilon _k}\left[ l \right],\forall k$.

\subsubsection{Wiretap Channels for IUs} Generally speaking, the BS does not know the channel information of the IUs, including their position and angle, but since the BS is able to detect potential IUs, it has knowledge of the LoS channel, which is often imprecise, including the distance and angle of the potential IUs. Meanwhile, due to the presence of scatterers in the environment and multipath fading, the NLoS channel of an IU is usually unknown, and in this paper it is assumed that its elements are bounded by an upper boundary, i.e., $\left|{\underline g}_{m,n}\left[l\right]\right|\le {\varepsilon _{m,n}\left[l\right]}$\footnote{In general, for the NLoS part of the Rician fading model, which is not bounded, however, it is possible to ensure that the constraints hold by selecting a sufficiently large ${\varepsilon _{m,n}\left[l\right]}$, which makes the probability of not satisfying the constraint small enough.}, where ${\underline g}_{m,n}\left[l\right]$ is the $n$-th element of ${\underline {\bf{g}}_m}\left[ l \right]$. For IUs, the channel uncertainties we consider include the distance between the IU and the BS, and the azimuth and pitch angles of the IU, so the wiretap channel ${{\bf{g}}_m\left[ l \right]} \in {\mathbb{C}^{N \times 1}}$ between a potential IU and the BS can be expressed as shown in Eq. (\ref{gm}).

 \begin{figure*}[ht] 
	\centering
	\begin{equation}
		{{\bf{g}}_m}\left[ l \right] = \sqrt {\frac{{{\lambda ^2_c}}}{{(4\pi)^2 \left( {1 + {\kappa _v}} \right){{\left( {{\widehat d_m} + \Delta {d_m}\left[ l \right]} \right)}^2}}}} \left[ {\sqrt {{\kappa _v}} {\left[ {{e^{ - j2\pi {{\overline  \delta} _r}{{\bf{n}}_r}}}} \right]^T} \otimes {\left[ {{e^{ - j2\pi {{\overline  \delta} _c}{{\bf{n}}_c}}}} \right]^T} + {\underline {\bf{g}}_m}\left[ l \right]} \right],\forall m,
		\label{gm}
	\end{equation}
	\hrulefill
\end{figure*}

{\noindent The $\Delta d_m\left[l\right]$ denotes the distance uncertainty and satisfies $\left| {\Delta {d_m}\left[ l \right]} \right| \le {D_m}\left[ l \right].$ Meanwhile, the ${\bar \delta _r}$ and ${\bar \delta _r}$ in Eq. (\ref{gm}) can be expressed as }

\begin{equation}
			\setlength{\abovedisplayskip}{3pt}
	\setlength{\belowdisplayskip}{3pt}
	{{\overline  \delta} _r}={d_f}\sin \left({\widehat\theta _m}+{\Delta \theta _m\left[ l \right]}\right)\cos \left({\widehat\varphi _m}+{\Delta \varphi _m\left[ l \right]}\right)/{\lambda _c},
\end{equation}
\begin{equation}
			\setlength{\abovedisplayskip}{3pt}
	\setlength{\belowdisplayskip}{3pt}
	{{\overline  \delta} _c}={d_f}\sin \left({\widehat\theta _m}+{\Delta \theta _m\left[ l \right]}\right)\sin \left({\widehat\varphi _m}+{\Delta \varphi _m\left[ l \right]}\right)/{\lambda _c},
\end{equation}
where $\left| {\Delta {\theta _m}\left[ l \right]} \right| \le {\Theta _m}\left[ l \right]$ and $\left| {\Delta {\varphi _m}\left[ l \right]} \right| \le {\Phi _m}\left[ l \right]$ denote the angle uncertainties. We assume that the errors of these parameters are bounded rather than derived, and the acquisition of angle and distance parameters in practice can be found in references\cite{Sadi2012,Yin2022a,Xu2006}. The uncertainty terms about $\Delta d_m\left[l\right]$, $\Delta \theta _m \left[l\right]$ and $\Delta \varphi _m\left[l\right]$ are tricky for the secure ISAC design, and in order to address thses problems, we handle them in Section IV. For ease of expression, we collect these uncertainties into the set 
\begin{equation}
	\begin{split}
			{\Xi _m}\left[ l \right] \buildrel \Delta \over = \left\{ {\left| {{\underline{g}_{m,n}}\left[ l \right]} \right| \le {\varepsilon _{m,n}}\left[ l \right],\left| {\Delta {d_m}\left[ l \right]} \right| \le {D_m}\left[ l \right],}\right.\\
			\left.	{\left| {\Delta {\theta _m}\left[ l \right]} \right| \le {\Theta _m}\left[ l \right],\left| {\Delta {\varphi _m}\left[ l \right]} \right| \le {\Phi _m}\left[ l \right]} \right\}.
	\end{split}
\end{equation}
\subsubsection{Radar Sensing Channels for IUs} The TRIS transceiver realizes the function of multiple antenna and its transmit steering vector with respect to ${\left( {{\theta _m}\left[l\right],{\varphi _m}\left[l\right]} \right)}$ can be represented as
\begin{equation}
	\setlength{\abovedisplayskip}{3pt}
	\setlength{\belowdisplayskip}{3pt}
	{\bf{a}}_m\left[l\right]={\left[ {{e^{ - j2\pi {{\overline  \delta} _r}{{\bf{n}}_r}}}} \right]^T} \otimes {\left[ {{e^{ - j2\pi {{\overline  \delta} _c}{{\bf{n}}_c}}}} \right]^T},\forall m.
\end{equation}
Since the TRIS transceiver is equipped with a feed antenna, its receive channel gain can be expressed as
\begin{equation}
	\setlength{\abovedisplayskip}{3pt}
	\setlength{\belowdisplayskip}{3pt}
	b_m\left[l\right] = {e^ {- j2\pi {d_m}/\lambda_c} },\forall m.
\end{equation}
Ultimately the sensing channel with respect to the IU $m$ can be expressed as
\begin{equation}
	{{\bf{c}}}_m\left[l\right]=b_m\left[l\right]{\bf{a}}^H_m\left[l\right],\forall m.
\end{equation}
\subsection{Signal Model}

In this paper, rate split signaling is considered where the transmission information is categorized into common and private streams. The signal can be expressed as

\begin{equation}
			\setlength{\abovedisplayskip}{3pt}
	\setlength{\belowdisplayskip}{3pt}
	\begin{split}
		{\bf{x}}\left[ l \right] &= {{\bf{W}}\left[ l \right]{\bf{s}}\left[ l \right]},\\
		& = \underbrace {{{\bf{w}}_c}\left[ l \right]{s_c}\left[ l \right]}_{{\rm{Interference~for~IU}}} + {\sum\nolimits_{i = 1}^K {{{\bf{w}}_i}\left[ l \right]{s_i}\left[ l \right]} },
	\end{split}
\end{equation}
where ${\bf{W}}\left[ l \right] = \left[ {{{\bf{w}}_c}\left[ l \right],{{\bf{w}}_1}\left[ l \right], {{\bf{w}}_2}\left[ l \right],\cdots ,{{\bf{w}}_k}}\left[ l \right] \right] \in \mathbb{C}{^{N \times \left( {K + 1} \right)}}$ denotes the linear beamforming matrix, and ${\bf{s}}\left[ l \right] = {\left[ {{s_c\left[ l \right]},{s_1\left[ l \right]}, {s_2\left[ l \right]}, \cdots ,{s_K\left[ l \right]}} \right]^T} \in \mathbb{C}{^{\left( {K + 1} \right) \times 1}}$ represents the transmission information, which is a wide stationary process in the time domain, statistically independent and with zero mean. In this paper, the common information of all LUs are combined and jointly encoded into a common stream $s_c$ using the codebook codes shared by all LUs. The private information of all LUs are encoded independently as private streams $\left\{{s_1},\cdots ,{s_K}\right\}$, which are only decoded by the corresponding user. For LU, the common stream is used for communication and is friendly because they can decode this part of the information. For IUs, the common stream is considered as AN and cannot be decoded.

In addition, it is reasonable and effective for the common stream to be used to interfere with the IUs due to the fact that prior to the scanning period, the LUs interact with the BS, which schedules the LUs, and the common stream encoding and decoding codebooks are shared between the LUs and the BS, and the IUs do not interact with the BS to obtain the codebook jointly encoded by the common stream, and therefore the IUs are unable to decode the information. 

The signal received by the $k$-th LU can be expressed as
\begin{equation}
	{y_k\left[l\right]} = \sum\nolimits_{i = 1}^{\cal \hat K}{\bf{h}}_k^H\left[l\right]{{{\bf{w}}_i}}\left[l\right]{{s_i}\left[l\right]}+ {n_k\left[l\right]},\forall k,
\end{equation}
where ${n_k\left[l\right]} \sim {\cal C}{\cal N}\left( {0,{\sigma _{{n_k}}^2}} \right)$ denotes the Gaussian white noise at the LU and ${\cal \hat K}=\left\{c,1,...,K\right\}$. 

The signal received by the $m$-th IU can be expressed as
\begin{equation}
			\setlength{\abovedisplayskip}{3pt}
	\setlength{\belowdisplayskip}{3pt}
	\begin{split}
			{y_m\left[l\right]} = \underbrace {{\bf{g}}_m^H\left[l\right]{{\bf{w}}_c}\left[ l \right]{s_c}\left[ l \right]}_{{\rm{AN}}}+\sum\limits_{i = 1}^K{\bf{g}}_m^H\left[l\right]{{{\bf{w}}_i}}\left[l\right]{{s_i}\left[l\right]}+ {v_m\left[l\right]},\forall m,
	\end{split}
\end{equation}
where ${v_m\left[l\right]} \sim {\cal C}{\cal N}\left( {0,{\sigma _{{v_m}}^2}} \right)$ denotes the Gaussian white noise at the IU.
\section{Quality of Service Evaluation Metrics And Problem Formulation}
In this section, we define the performance metrics for secure communication and sensing, respectively. And formulate the robust secure ISAC resource allocation optimization problem.
\subsection{Secure Communication Performance Metric}
In this paper, we consider system secrecy sum-rate and outage probabilities as performance criteria for secure communication. The secrecy rate is designed to meet the rate requirements of LUs, and the secrecy outage probability is designed to minimize eavesdropping by IUs.

For the secrecy rate, due to the utilization of rate-spliting signaling, when LUs decoding the common stream, the private stream is considered as interference. After decoding the common stream, this part is removed in the received signal and the remaining private stream is considered as interference for decoding the current LU's information. Then, the corresponding common and private stream signal-to-noise ratio (SNR) of the $k$-th LU can be expressed as follows

\begin{equation}
			\setlength{\abovedisplayskip}{3pt}
	\setlength{\belowdisplayskip}{3pt}
	{\gamma _{c,k}}\left[l\right] = \frac{{{{\left| {{\bf{h}}_k^H\left[l\right]{{\bf{w}}_{c}}\left[l\right]} \right|}^2}}}{{\sum\nolimits_{i = 1}^K {{{\left| {{\bf{h}}_k^H\left[l\right]{{\bf{w}}_{i}}\left[l\right]} \right|}^2}}  + {\sigma _{{n_k}}^2}}}, \forall k,
\end{equation}
and
\begin{equation}
	{\gamma _{p,k}}\left[l\right] = \frac{{{{\left| {{\bf{h}}_k^H\left[l\right]{{\bf{w}}_{k}}\left[l\right]} \right|}^2}}}{{\sum\nolimits_{i \ne k}^K {{{\left| {{\bf{h}}_k^H\left[l\right]{{\bf{w}}_{i}}\left[l\right]} \right|}^2}}  + {\sigma _{{n_k}}^2}}}, \forall k,
\end{equation}
Then the corresponding achievable rate can be expressed as
\begin{equation}
	{R_{i,k}}\left[l\right] = {\log _2}\left( {1 + {\gamma _{i,k}}\left[l\right]} \right), i \in \left\{{c,p}\right\}, \forall k.
\end{equation}
As mentioned earlier, the common stream is shared by all LUs and jointly participates in the encoding of the information. In order to ensure that all LUs are able to decode the common stream, the following constraints need to be satisfied
\begin{equation}
	{R_c}\left[l\right] = \min \left( {{R_{c,1}}\left[l\right], \cdots ,{R_{c,K}}}\left[l\right] \right), 
\end{equation}
and
\begin{equation}
\sum\nolimits_{k=1}^K {{C_k}\left[l\right]}  = {R_c\left[l\right]},
\end{equation}
where ${C_k\left[l\right]}$ denotes the equivalent common stream rate of the $k$-th LU. Let ${\bf c}\left[l\right]=\left\{C_1\left[l\right],\cdots,C_K\left[l\right]\right\}$ denotes the common stream vector. Then the achievable rate for the $k$-th LU can be expressed as

\begin{equation}
	R_k\left[l\right] = C_k\left[l\right] + R_{p,k}\left[l\right], \forall k.
\end{equation} 

In this paper we consider the worst-case scenario where an IU can eliminate the private stream interference from the remaining LUs before decoding the information of a specific LU\footnote{We consider that the IUs have enough computational power to decode the private stream interference of the remaining LUs when eavesdropping on a specific LU, but cannot decode the common stream information since decoding the common stream requires interaction with the BS. If an illegal user interacts with the BS it will be exposed, resulting in a communication outage.}. As mentioned earlier, since the IU cannot decode the common stream as AN, in the time slot $l$, the eavesdropping signal-to-noise ratio of the $m$-th IU to the $k$-th LU can be expressed as

\begin{equation}
				\setlength{\abovedisplayskip}{3pt}
	\setlength{\belowdisplayskip}{3pt}
		{\gamma _{k,m}}\left[l\right] = \frac{{{{\left| { {{\bf{g}}_m^H}\left[l\right] {{\bf{w}}_k}}\left[l\right] \right|}^2}}}{{ {{{\left| {{{\bf{g}}_m^H}\left[l\right] {{\bf{w}}_{c}}\left[l\right]} \right|}^2}}  + {\sigma _{{v_m}}^2}}}, \forall k,\forall m,
\end{equation}
Then the corresponding eavesdropping rate of the $m$-th IU on the $k$-th LU can be expressed as
\begin{equation}
				\setlength{\abovedisplayskip}{3pt}
	\setlength{\belowdisplayskip}{3pt}
	{R_{k,m}}\left[l\right] = {\log _2}\left( {1 + {\gamma _{k,m}}\left[l\right]} \right), \forall k, \forall m,	
\end{equation}
Therefore, the system secrecy sum-rate in time slot $l$ can be expressed as
\begin{equation}
				\setlength{\abovedisplayskip}{3pt}
	\setlength{\belowdisplayskip}{3pt}
	R_{tot}^{secure}\left[l\right] = \sum\nolimits_{k=1}^K\left[R_{k}\left[l\right] - \mathop{\rm {max}}\limits_{m \in M} R_{k,m}\left[l\right]\right]^+,\forall l,
\end{equation}
where $\left[x\right]^+$ denotes ${\rm max}\left\{0,x\right\}$.

For the outage probability, this paper considers two cases. The first case is considered on the LU side, which can be written in the following form

\begin{equation}
				\setlength{\abovedisplayskip}{3pt}
	\setlength{\belowdisplayskip}{3pt}
	\Pr \left\{ {{R_k}\left[l\right] \ge r_k^s} \right\} \ge 1 - {P_{out,1}},\forall k,
\end{equation}
This rate outage constraint emphasizes service reliability, i.e., it guarantees that LU $k$ can still decode rate $r_k^s$ with probability $1 - {P_{out,1}}$ under the channel state information error. The second case is considered on the IU side and can be written in the following form
\begin{equation}
				\setlength{\abovedisplayskip}{3pt}
	\setlength{\belowdisplayskip}{3pt}
	\Pr \left\{ {{R_{k,m}}\left[l\right] \le r_m^e} \right\} \ge 1-{P_{out,2}},\forall k, m,
\end{equation}
This constraint ensures that the probability that the eavesdropping rate of an IU is less than a certain threshold $r_m^e$ is not less than $1-{P_{out,2}}$.
\subsection{Sensing Performance Metric}
In general, evaluations of ISAC systems can be categorized into two types, namely, information-based metrics and estimation-based metrics. Information-based metrics include radar mutual information, etc. Estimation-based metrics include CRB, mean square error (MSE) and detection probability, etc. As mentioned above, the BS needs to constantly detect potential LUs during beam scanning, so the detection probability is naturally one of the metrics for sensing. In addition, a well-designed waveform is required to achieve better communication and sensing trade-offs as well as to realize high-precision sensing. Therefore, we adopt the detection probability and beampattern approximation as the performance metrics in this paper.
\subsubsection{Detection Probability} In this paper, likelihood ratio detection\cite{Trees2001} is applied, i.e., a threshold $\delta$ is set, when $T$ is greater than this threshold the target is considered to exist and recorded as ${\cal H}_1$, and vice versa as ${\cal H}_0$. $T$ obeys the following distribution
\begin{equation}
	T \sim \left\{ {\begin{array}{*{20}{l}}
			{\frac{{\sigma _r^2}}{2}{\cal X}_{\left( 2 \right)}^2,}&{{{\cal H}_0}}\\
			{\left( {\frac{{\sigma _r^2}}{2} + \frac{{\sigma _{{\alpha _m}}^2{ {{{{p}}}\left[l\right]} }{{\left\| {{\bf{c}}}_m\left[l\right] \right\|}^2}}}{2}} \right){\cal X}_{\left( 2 \right)}^2,}&{{{\cal H}_1}}
	\end{array}}, \right.
\end{equation}
where ${\cal X}_{\left( 2 \right)}^2$ denotes the Chi-square distribution with degree of freedom 2, $p\left[l\right]={\rm {tr}}\left({{\bf{w}}_c}\left[l\right]{{\bf{w}}_c^H}\left[l\right]+\sum_{k = 1}^K{{\bf{w}}_k}\left[l\right]{{\bf{w}}_k^H}\left[l\right]\right)$, and ${\sigma _r^2}$ denotes processing noise. ${\alpha _m} \sim {\cal C}{\cal N}\left( {0,\sigma _{{\alpha _m}}^2} \right)$ denotes the complex reflection coefficient, $\sigma _{{\alpha _m}}^2 = {S_{RCS}} {\lambda_c}^2 / {((4\pi)^3d_m^4)}$, and ${S_{RCS}}$ denotes the RCS of target. 

With the above likelihood detection test defined, the false alarm probability can be expressed as
\begin{equation}
	\begin{split}
		{P_{FA}} = \Pr \left( {T > \delta |{{\cal H}_0}} \right) &= \Pr \left( {\frac{{\sigma _r^2}}{2}{\cal X}_{\left( 2 \right)}^2 > \delta } \right) \\&= \Pr \left( {{\cal X}_{\left( 2 \right)}^2 > \frac{{2\delta }}{{\sigma _r^2}}} \right),
	\end{split}
\end{equation}
Then given a $P_{FA}$, the threshold can be expressed as
\begin{equation}
	\delta  = \frac{{\sigma _r^2}}{2}F_{{\cal X}_{\left( 2 \right)}^2}^{ - 1}\left( {1 - {P_{FA}}} \right).
\end{equation}
where ${F_{{\cal X}_{\left( 2 \right)}^2}}^{-1}$  denotes the inverse of the cumulative distribution function of the chi-square distribution. Finally, the probability of detection of a potentially IU $m$ in the current time slot $l$ can be expressed in the following form

\begin{equation}
				\setlength{\abovedisplayskip}{3pt}
	\setlength{\belowdisplayskip}{3pt}
	\begin{split}
	{P_m^{detection}}\left[l\right] &= \Pr \left( {T > \delta |{{\cal H}_1}} \right)\\
	&=1 - {F_{{\cal X}_{\left( 2 \right)}^2}}\left( {\frac{{{{2\delta } \mathord{\left/
						{\vphantom {{2\delta } {\sigma _r^2}}} \right.
						\kern-\nulldelimiterspace} {\sigma _r^2}}}}{{1 + p\left[l\right]{\varsigma _m\left[l\right]}}}} \right), \forall l,
	\end{split}
\end{equation}
where ${F_{{\cal X}_{\left( 2 \right)}^2}}$  denotes the cumulative distribution function of the chi-square distribution. The normalized sensing channel gain $\varsigma _m\left[l\right]$ can be expressed as\cite{9945983,10522473}

\begin{equation}
				\setlength{\abovedisplayskip}{3pt}
	\setlength{\belowdisplayskip}{3pt}
	{\varsigma _m}\left[l\right] ={\sigma _{{\alpha _m}}^2{ \left|{{\bf{c}}}_m\left[l\right]{\widehat{\bf{c}}^H_m\left[l\right]} \right|^2}}/{N^2{\sigma _r^2}},\forall l,
\end{equation}
where  ${\widehat{\bf{c}}}_m\left[l\right]={\widehat b}_m\left[l\right]{\bf{\widehat a}}^H_m\left[l\right]$  denotes the estimated radar channel with respect to ${\left( {{\widehat\theta _m}\left[l\right],{\widehat\varphi _m}\left[l\right]} \right)}$ and $\widehat d_m$. 

\subsubsection{Beampattern Approximation} Communication is the utilization of random signals for transmission, while sensing is the detection of targets and acquisition of parameters in a deterministic wireless environment, and the difference between the two will pose a challenge to ISAC signal design. To address this challenge, we adopt a beampattern approximation scheme, i.e., we differentiate a random signal from a pre-designed signal with high directionality. This integrated waveform design satisfies the random characteristics of communication signals and ensures the accuracy of detection of potential IUs/targets in a specific direction. Therefore, the waveform design at timeslot $l$ needs to satisfy the following constraints.

\begin{equation}
			\setlength{\abovedisplayskip}{3pt}
	\setlength{\belowdisplayskip}{3pt}
	\left\| {\sum\nolimits_{i = 1}^{\hat {\cal K}} {{{\bf{W}}_i}\left[ l \right]}  - {\bf{R}}\left[ l \right]} \right\|_F^2 \le \delta \left[ l \right],\forall l,
\end{equation}
where ${\bf{R}}\left[ l \right]$ denotes the covariance matrix of the desired waveform\cite{1399140} and $\delta \left[ l \right]$ denotes the pre-designed error between the actual signal transmitted and the pre-designed high directional beampattern.
\subsection{Problem Formulation}

In this present paper, an optimization problem is constructed to maximize the secrecy  sum-rate as the objective function while guaranteeing the sensing performance as the constraints over $L$ time slots in the scanning period $T$. First we define ${\bf W}_k\left[ l \right]={\bf w}_k\left[ l \right]{\bf w}_k^H\left[ l \right]$ and ${\bf W}_c\left[ l \right]={\bf w}_c\left[ l \right]{\bf w}_c^H\left[ l \right]$, then design the optimization variables $\left\{t\left[ l \right], C_k\left[ l \right], {\bf W}_k\left[ l \right], {\bf W}_c\left[ l \right]\right\}$ by solving the following optimization problem
\begin{subequations}
	\begin{align}
		&\left( {{\textrm{P0}}} \right){\rm{:~}}{\mathop \textrm{max}\limits_{t\left[ l \right],{C_k}\left[ l \right],{{\bf{W}}_k}\left[ l \right],{{\bf{W}}_c}\left[ l \right]}}{\widetilde R_{tot}^{secure}}, \notag\\
		&~~{\rm{s}}{\rm{.t}}{\rm{.}}~~~{\rm{rank}}\left( {{{\bf{W}}_i}\left[ l \right]} \right) = 1,\forall i \in \hat {\cal K},\forall l,\label{rank1}\\
		&~~~~~~~~~{{\bf{W}}_i}\left[ l \right] \succeq 0,\forall i \in \hat {\cal K},\forall l,\label{suc}\\
		&~~~~~~~~~{\sum\nolimits_{i = 1}^{\hat {\cal K}} {\left[ {{{\bf{W}}_i}\left[ l \right]} \right]} _{nn}} \le {P_t},\forall n, l,\\
		&~~~~~~~~~\sum\nolimits_{l = 1}^L {t\left[ l \right]}  \le T,\\
		&~~~~~~~~~{t_{\min }} \le t\left[ l \right] \le {t_{\max }},\forall l,\label{tli}\\
		&~~~~~~~~~{C_k}\left[ l \right] \ge 0,\forall k, l,\label{Ck}\\
		&~~~~~~~~~\sum\nolimits_{k = 1}^K {{C_k}\left[ l \right]}  \le {R_c}\left[ l \right],\forall l,\label{C0}\\
		&~~~~~~~~~\Pr \left\{ {{R_k}\left[ l \right] \ge r_k^s} \right\} \ge 1 - {P_{out,1}},\forall k, l,\label{P1}\\
		&~~~~~~~~~\Pr \left\{ {{R_{k,m}}\left[ l \right] \le r_m^e} \right\} \ge 1 - {P_{out,2}},\forall k, m, l,\label{P2}\\
		&~~~~~~~~~P_m^{detection}\left[ l \right] \ge {P_D},\forall m, l,\label{PD}\\
		&~~~~~~~~~\left\| {\sum\nolimits_{i = 1}^{\widehat K} {{{\bf{W}}_i}\left[ l \right]}  - {\bf{R}}\left[ l \right]} \right\|_F^2 \le \delta \left[ l \right],\forall l.\label{beam}
	\end{align}	
\end{subequations}
Since space is limited, the objective function ${\widetilde R_{tot}^{secure}}$ is represented as shown in Eq. (\ref{Rse}). It can be seen that problem (P0) is a nonconvex problem, due to the coupling ofvariables and nonconvex operations. In next section, we will discuss how to transform it into a convex problem and solve it. For channel errors, this paper employs both bounded and probabilistic models, which are compatible under certain settings. Regarding the bounded error modeling, we apply it to the objective function due to the fact that the active and passive estimation of the channel in this paper results in channel errors that are often bounded, and that the objective function is a function of the channel errors. Regarding the probabilistic error model, we apply it to the constraints and consider it from the point of view that the error is required to obey a Gaussian distribution since the outage probability is used as a constraint. In concrete terms, we can set a sufficiently large upper bound on the error, so that the probability that the error exceeds this upper bound is sufficiently small, in such a way that the two models are compatible.
 \begin{figure*}[ht] 
	\centering
	\begin{equation}
		{\widetilde R_{tot}^{secure}}={\frac{1}{T}}\sum\nolimits_{l = 1}^L {t\left[ l \right]}\sum\nolimits_{k = 1}^K {{{\left[ {\mathop {\min }\limits_{\Delta {{\bf{h}}_k}\left[ l \right]} {R_k}\left[ l \right]} - {\mathop {{\rm{max}}}\limits_{m \in M} {\mathop {\max }\limits_{{\Xi _m}\left[ l \right]} {R_{k,m}}\left[ l \right]}} \right]}^ + }}.\label{Rse}
	\end{equation}
	\hrulefill
\end{figure*}

\section{Joint Design of Communication and Sensing}
In this section, we first give the bound for uncertainty region of IUs' channel, and then the above non-convex problem (P0) is transformed into a convex problem and solved, along with the joint design of communication and sensing parameters, and finally the complexity and convergence analysis of the algorithm is given.
\begin{figure}[H]
	\centerline{\includegraphics[width=9cm]{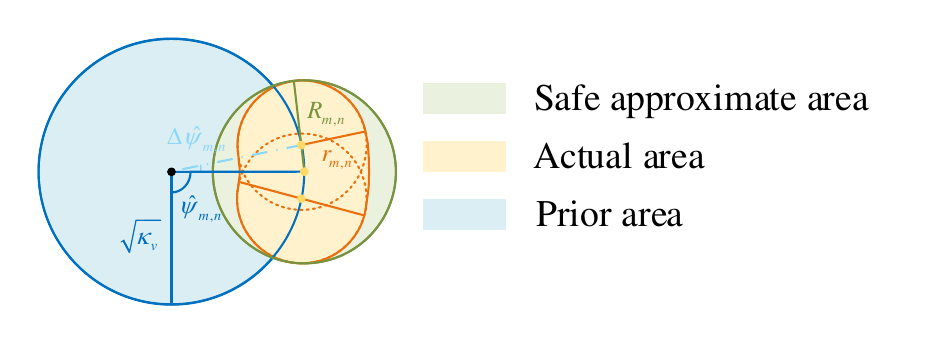}}
	\caption{Small-scale fading uncertainty.}\label{bound}
\end{figure}
\subsection{Bound for IUs' Channel Error}
Firstly, we consider the angle and the small-scale fading uncertainties of IUs' channel. Regarding the distance error associated with large-scale fading, we address in the subsequent. The Eq. (\ref{gm}) can be rewrited as

\begin{equation}
			\setlength{\abovedisplayskip}{3pt}
	\setlength{\belowdisplayskip}{3pt}
\begin{split}
	{{\bf{g}}_m}\left[ l \right] &= \sqrt {\frac{{\xi _m^2\left[ l \right]}}{{\left( {1 + {\kappa _v}} \right)}}} \left[ {\sqrt {{\kappa _v}}  + \Delta {g_{m,1}}\left[ l \right],} \right.\\
	&~~~~~~~~~~~\left. { \cdots ,\sqrt {{\kappa _v}} {e^{ - j {\widehat\psi _{m,N}}\left[l\right]}} + \Delta {g_{m,N}}\left[ l \right]} \right]^T,\end{split}\label{gmm}
\end{equation}
where $\xi_m\left[l\right]={{\lambda _c}}/{{4\pi {d_m}\left[ l \right]}}$, ${d_m}\left[ l \right]={{\widehat d_m} + \Delta {d_m}\left[ l \right]}$, and $\Delta g_{m,n}$ contains the combined effects of angle and small-scale fading uncertainties. As shown in Fig. 4, $\sqrt {{\kappa _v}}$ denotes the radius of the blue circle and $\Delta {g_{m,n}}\left[ l \right]$ denotes the radius of the orange circle, and the values of the channel elements are the points on the orange circle without considering large-scale fading. For known estimation of the channel we denote the a priori region in blue, for the actual uncertainty region we denote it in yellow, and for the safe approximation we denote it in green.

Existing work considers uncertainties in azimuth and small-scale fading under uniform linear array\cite{9933849}, and the analysis in this paper considers uncertainties in azimuth and pitch angle as well as small-scale fading under the UPA. Let $n=N_cn_r+n_c+1$ , $\widehat\psi_{m,n}\left[l\right]$ in Eq. (\ref{gmm}) can be expressed as follow
\begin{equation}
				\setlength{\abovedisplayskip}{3pt}
	\setlength{\belowdisplayskip}{3pt}
	\widehat\psi_{m,n}\left[l\right]={2\pi }({{\widehat  \delta} _r}{{{n}}_r}+{{\widehat  \delta} _c}{{{n}}_c}),\forall m,n.
\end{equation}
where ${{\widehat  \delta} _r}$ and ${{\widehat  \delta} _c}$ are trigonometric functions with respect to ${\widehat\theta _m}$ and ${\widehat\varphi _m}$, similarly defined as Eqs. (8) and (9).  
Thus the $\Delta \psi_{m,n}\left[l\right]$ is given by
\begin{equation}
				\setlength{\abovedisplayskip}{3pt}
	\setlength{\belowdisplayskip}{3pt}
	\begin{split}
	\Delta {\psi _{m,n}}\left[ l \right] = {{2\pi }}\left|{{\widehat  \delta} _r}{{{n}}_r}+{{\widehat  \delta} _c}{{{n}}_c}-{{\overline  \delta} _r}{{{n}}_r}-{{\overline  \delta} _c}{{{n}}_c}  \right|,\forall m,n.
	\end{split}
\end{equation}
According to geometric theory, the actual uncertainty region consists of two semicircles and a segment of a ring, so its area can be expressed as
\begin{equation}
	\begin{split}
			S_{act} =\pi r_{m,n}^{2}+4\sqrt {{\kappa _v}}r_{m,n}{\Delta {\psi _{m,n}}\left[ l \right]}/{\pi},\forall m,n.
	\end{split}
\end{equation}
By the cosine theorem, we have
\begin{equation}
	R_{m,n} = r_{m,n}+\sqrt {2{\kappa _v}}\sqrt{1-\cos{\Delta {\psi _{m,n}}\left[ l \right]} },\forall m,n.
\end{equation}
Then the area of the safe approximation region can be expressed as
\begin{equation}
	S_{saf} = \pi\left(r_{m,n}+\sqrt {2{\kappa _v}}\sqrt{1-\cos{\Delta {\psi _{m,n}}\left[ l \right]} }\right)^2,\forall m,n.
\end{equation}
The ratio of $S_{act}$ and $S_{saf}$ indicates the accuracy of the approximation, which was verified in \cite{9933849}, and the approximation is accurate when the angular uncertainty is small. Thus, the uuper bound of the uncertainty term $\Delta g_{m,n}\left[l\right]$ can be expressed as
\begin{equation}
				\setlength{\abovedisplayskip}{3pt}
	\setlength{\belowdisplayskip}{3pt}
	\left|\Delta g_{m,n}\left[l\right]\right|\le {\varepsilon _{m,n}\left[l\right]}+\sqrt{2\kappa_v}\sqrt{1-\cos\left(\Delta\psi_{m,n}\left[l\right]\right)}.
\end{equation}
Then, we collect all the $\Delta g_{m,n}\left[l\right]$ as $\Delta {\bf g}_{m}\left[l\right]\in \mathbb{C}^{N\times1}$. Therefore, the upper bound on uncertainty for the $m$-th IU $\Delta {\bf g}_{m}\left[l\right]$ can be expressed as
\begin{equation}
				\setlength{\abovedisplayskip}{3pt}
	\setlength{\belowdisplayskip}{3pt}
	\left\| {\Delta {\bf g}_{m}\left[l\right]} \right\|\le \tau_m\left[l\right],\forall m,
\end{equation}
where $\tau_m^2\left[l\right]=\sum\nolimits_{n = 1}^N\left({\varepsilon _{m,n}\left[l\right]}+\sqrt{2\kappa_v}\sqrt{1-\cos\left(\Delta\psi_{m,n}\left[l\right]\right)}\right)^2$.
Similarly, we assume the channel uncertainty for IUs obeys the distribution of CSCG, i.e., $\Delta {{\bf{g}}_m}\left[ l \right]\sim {\cal C}{\cal N}\left( {0,{{\bf{E}}_{e,m}}} \right)$, ${{\bf{E}}_{e,m}}\succ 0$. This is reasonable because when $\Delta {{\bf{g}}_m}\left[ l \right]$ is bounded, then the variance of its distribution will be able to be given. Then, we rewrite the channel for IUs as follow
\begin{equation}
	{\bf g}_m\left[l\right] = \sqrt{\frac{\xi_m^2\left[l\right]}{1+\kappa_v}}\left({\widehat{\bf g}}_m+{\Delta {\bf g}_{m}\left[l\right]}\right),\forall m,
\end{equation}
where ${\widehat{\bf g}}_m=\left[\sqrt{\kappa_v},\cdots,\sqrt{\kappa_v}e^{-j\widehat\psi_{m,N}\left[l\right] }\right]^T$ denotes the estimate channel for the potential $m$-th IU before the scanning period $T$. 

\subsection{Convex Transformation of the Formulated Problem} 
\subsubsection{Common Rate Constraint}We first deal with constraints. It can be seen that constraint (\ref{C0}) is not a non-convex set, which can be transformed to

\begin{equation}
			\setlength{\abovedisplayskip}{3pt}
	\setlength{\belowdisplayskip}{3pt}
\begin{split}
	&{\rm{tr}}\left( {{{\bf{h}}_k}\left[ l \right]{\bf{h}}_k^H\left[ l \right]{{\bf{W}}_c}\left[ l \right]} \right) \\
	&\ge {\gamma _{c0}}\left( {\sum\nolimits_{i = 1}^K {{\rm{tr}}\left( {{{\bf{h}}_k}\left[ l \right]{\bf{h}}_k^H\left[ l \right]{{\bf{W}}_i}\left[ l \right]} \right)}  + {\sigma _{n_k}^2}} \right),\forall k,l,\label{co1}
\end{split}
\end{equation}
where ${\gamma _{c0}} = {2^{{R_c}\left[ l \right]}} - 1$ denotes the signal-to-noise ratio corresponding to $\sum\nolimits_{k = 1}^K {{C_k}\left[ l \right]} $. Notice that $\mathbb{E}\left\{{\bf h}_k\left[ l \right]{\bf h}_k^H\left[ l \right]\right\}=\widehat{\bf h}_k\left[ l \right]\widehat{\bf h}_k^H\left[ l \right]+{\varepsilon _k^2}\left[ l \right]{\bf I}_N \buildrel \Delta \over = {\bf H}_k\left[l\right]$\footnote{Since $\widehat{\bf h}_k$ and $\Delta {{\bf{h}}_k}\left[ l \right]$ are independent, we have $\mathbb{E}\left\{\widehat{\bf h}_k^H\Delta {{\bf{h}}_k}\left[ l \right]\right\}=0$, and assuming that the channel is constant at each transmission timeslot $l$, we can ignore the operation of the channel covariance expectation operator.}, so Eq. (\ref{co1}) is expressed as
\begin{equation}
	\begin{split}
		&{\rm tr}\left({\bf H}_k\left[l\right]{\bf W}_c\left[l\right]\right)\\
		&~~~~~~~\ge {\gamma _{c0}}\left( {\sum\nolimits_{k = 1}^K {{\rm{tr}}\left( {{\bf H}_k\left[l\right]{{\bf{W}}_k}\left[ l \right]} \right)}  + {\sigma _{n_k}^2}} \right),\forall k,l.\label{co}
	\end{split}
\end{equation}

\subsubsection{Secure Outage Constraint}For constrain (\ref{P1}), we first treat it in the following form

\begin{equation}
	\Pr \left\{{\gamma _{p,k}}\left[l\right]\le 2^{{r_k^s-C_k\left[l\right]}}-1\right\}\le P_{out,1},
\end{equation}
Then, making variable substitutions and simplifications, there are
\begin{equation}
	\begin{split}
			&\Pr \left\{ {\Delta {{\bf{h}}_k}{{\left[ l \right]}^H}{{\overline {\bf{W}} }_k}\left[ l \right]\Delta {{\bf{h}}_k}\left[ l \right]} \right.\\
			&~~~~~~~\left. {  +2{\rm{Re}}\left\{ {\Delta {{\bf{h}}_k}{{\left[ l \right]}^H}{{\overline {\bf{W}} }_k}\left[ l \right]{{\widehat {\bf{h}}}_k}} \right\} \le \sigma _{{h,k}}^2} \right\} \le {P_{out,1}},\label{qua}
	\end{split}
\end{equation}
where ${{\overline {\bf{W}} }_k}\left[ l \right]=\frac{1}{2^{{r_k^s-C_k\left[l\right]}}-1}{\bf{W}}_k\left[ l \right]-\sum\nolimits_{i \ne k}^K{\bf{W}}_i\left[ l \right]$ and $\sigma _{{h,k}}^2=\sigma _{{n_k}}^2-{{\widehat {\bf{h}}}_k^H}{{\overline {\bf{W}} }_k}{{\widehat {\bf{h}}}_k}$. To normalize the Gaussian variable, we introduce the normalization factors as follows
\begin{equation}
	\begin{split}
		&\Pr \left\{ {{{\bf{p}}_k}{\left[ l \right]}^H}{{{ {\bf{A}} }_k}\left[ l \right] {{\bf{p}}_k}\left[ l \right]} \right.\\
		&~~~~~~~\left. {  +2{\rm{Re}}\left\{ { {{\bf{p}}_k}{\left[ l \right]}^H}{{{ {\bf{A}} }_k}\left[ l \right]{{\widehat {\bf{h}}}_k}} \right\} \le \sigma _{{h,k}}^2} \right\} \le {P_{out,1}},\label{qup}
	\end{split}
\end{equation}
where ${\bf p}_k\left[l\right]=\frac{1}{\sigma_{h_e}}\Delta {{\bf{h}}_k}{{\left[ l \right]}}$, ${\bf A}_k\left[l\right]={\sigma_{h_e}^2}{{\overline {\bf{W}} }_k}\left[ l \right]$ and ${\bf p}_k\left[l\right] \sim {\cal C}{\cal N}\left( {0,{{\bf{I}}_N}} \right)$.
Now, we have the standard form of quadratic, and utilize the Bernstien-type inequality for quadratic Gaussian proces as {\bf Lemma 1}. 
\begin{lemma}
	Bernstein-type inequalities{\rm \cite{bechar2009bernsteintype}:} Assume that the Gaussian variable ${\bf e} \sim {\cal C}{\cal N}\left( {0,{{\bf{I}}_N}} \right)$, the matrix ${\bf Q} \in \mathbb{H}^{N\times N}$ and ${\bf r}\in\mathbb{C}^{N\times1}$ satisfies $f({\bf e})={\bf e}^H{\bf Q}{\bf e}+2{\rm Re}\left\{{\bf e}^H{\bf r}\right\}$. Then for any nonnegative $\sigma$, we have
	\[\Pr \left\{ {f\left( {\bf{e}} \right) \ge E^+} \right\} \le {e^{ - \sigma }},\]
	and
	\[\Pr \left\{ {f\left( {\bf{e}} \right) \le E^-} \right\} \le {e^{ - \sigma }},\]
		where $E^+={\rm{tr}}\left( {\bf{Q}} \right) + \sqrt {2\sigma \left( {\left\| {\bf{Q}} \right\|_F^2 + 2{{\left\| {\bf{r}} \right\|}^2}} \right)}  + \sigma {\lambda^ + }\left( {\bf{Q}} \right)$, $E^-={\rm{tr}}\left( {\bf{Q}} \right) - \sqrt {2\sigma \left( {\left\| {\bf{Q}} \right\|_F^2 + 2{{\left\| {\bf{r}} \right\|}^2}} \right)}  - \sigma {\lambda^ - }\left( {\bf{Q}} \right)$, $\lambda^ +\left( {\bf{Q}} \right)={\rm max}\left\{\lambda_{max}({\bf Q}),0\right\}$, and $\lambda^ -\left( {\bf{Q}} \right)={\rm max}\left\{\lambda_{max}({-\bf Q}),0\right\}$.
\end{lemma}
If we can find a function $g$ such that $\Pr \left\{f\left( {\bf{e}} \right)\le 0\right\}\le g\left( {\bf{Q},{\bf r}} \right)$, then there are $g\left( {\bf{Q},{\bf r}} \right)\le P_{out} \Rightarrow \Pr \left\{f\left( {\bf{e}} \right)\le 0\right\}\le P_{out}$. Next, based on {\bf Lemma 1}, Eq. (\ref{qup}) is transformed into
\begin{equation}
	\begin{split}
	{\rm{tr}}\left( {{ {\bf{A}} }_k}\left[ l \right] \right) - \sqrt {2{\sigma}_1  }\sqrt { \left( {\left\| {{ {\bf{A}} }_k}\left[ l \right] \right\|_F^2 + 2{{\left\| {{ {\bf{A}} }_k}\left[ l \right]{{\widehat {\bf{h}}}_k} \right\|}^2}} \right)}\\
	  - {\sigma}_1 \lambda^-\left({\bf A}_k\left[l\right]\right)
	\ge \sigma _{{h,k}}^2,\forall k,\label{t1}
	\end{split}
\end{equation}
For solving the non-convexity, the Eq. (\ref{t1}) can eventually be transformed into \cite{6891348}
\begin{equation}
		{\rm{tr}}\left( {{ {\bf{A}} }_k}\left[ l \right] \right) - \sqrt {2{\sigma}_1  }z_k\left[l\right]  - {\sigma}_1 \nu_k\left[l\right]
		 \ge \sigma _{{h,k}}^2,\forall k,\label{qut}
\end{equation}
\begin{equation}
		 \left\|\left[ { vec\left({{ {\bf{A}} }_k}\left[ l \right]\right)  ; \sqrt{2}{{ {{ {\bf{A}} }_k}\left[ l \right]{{\widehat {\bf{h}}}_k} }}} \right]\right\|  \le z_k\left[l\right],\forall k,\label{soc}
\end{equation}
\begin{equation}
\nu_k\left[l\right]{\bf I}_N+{{ {\bf{A}} }_k}\left[ l \right]\succeq 0,\forall k,\label{el}		
\end{equation}
where ${\sigma}_1=-\ln\left(P_{out,1}\right)$, $z_k\left[l\right]$ denotes the non-negative slack variable, and then the second-order cone constraints $\sqrt { \left( {\left\| {{ {\bf{A}} }_k}\left[ l \right] \right\|_F^2 + 2{{\left\| {{ {\bf{A}} }_k}\left[ l \right]{{\widehat {\bf{h}}}_k} \right\|}^2}} \right)}$ are equivalent to (\ref{soc}). $vec\left(\cdot\right)$ denotes the transformation of a matrix into a column vector. Due to the operation $\lambda^-\left({\bf A}_k\left[l\right]\right)$ is nonconvex, we introduce the non-negative slack variable $\nu_k\left[l\right]$ to upper bound the maximum eigenvalue of $-{{ {\bf{A}} }_k}\left[ l \right]$ as (\ref{el}).

Similarly, for the constrain (\ref{P2}), it can be transformed into

\begin{equation}
			\setlength{\abovedisplayskip}{3pt}
	\setlength{\belowdisplayskip}{3pt}
	\Pr \left\{  {\gamma}_{k,m} \left[l\right]\ge 2^{r_{m}^e}-1\right\}\le P_{out,2},
\end{equation}
Making the variable substitutions and simplifications, there are
\begin{equation}
	\begin{split}
		&\Pr \left\{ {\Delta {{\bf{g}}_m}{{\left[ l \right]}^H}{{\underline {\bf{W}} }_k}\left[ l \right]\Delta {{\bf{g}}_m}\left[ l \right]} \right.\\
		&~\left. {  +2{\rm{Re}}\left\{ {\Delta {{\bf{g}}_m}{{\left[ l \right]}^H}{{\underline {\bf{W}} }_k}\left[ l \right]{{\widehat {\bf{g}}}_m}} \right\}+\sigma _{{g,m}}^2 \ge 0} \right\} \le {P_{out,2}},\label{qub}
	\end{split}
\end{equation}
where ${{\underline {\bf{W}} }_k}\left[ l \right]=\frac{1}{2^{r_m^e}-1}{\bf{W}}_k\left[ l \right]-{\bf{W}}_c\left[ l \right]$ and $\sigma _{{g,m}}^2={{\widehat {\bf{g}}}_m}^H{{\underline {\bf{W}} }_k}{{\widehat {\bf{g}}}_m}-\sigma _{{v_m}}^2$. To normalize the Gaussian variable, we introduce the normalization factors as follows
\begin{equation}
	\begin{split}
		&\Pr \left\{ {{{\bf{q}}_m}{\left[ l \right]}^H}{{{ {\bf{B}} }_m}\left[ l \right] {{\bf{q}}_m}\left[ l \right]} \right.\\
		&~~~\left. {  +2{\rm{Re}}\left\{ { {{\bf{q}}_m}{\left[ l \right]}^H}{{{ {\bf{B}} }_m}\left[ l \right]{{\widehat {\bf{g}}}_m}} \right\}+\sigma _{{g,m}}^2} \ge 0 \right\} \le {P_{out,2}},
	\end{split}
\end{equation}
where ${\bf q}_m\left[l\right]= \left({\bf E}_{e,m}^{1/2}\right)^{-1}{{\bf{g}}_m}{{\left[ l \right]}}$, ${\bf B}_m\left[l\right]={\bf E}_{e,m}^{1/2}{{\underline {\bf{W}} }_k}\left[ l \right]{\bf E}_{e,m}^{1/2}$ and ${\bf q}_m\left[l\right] \sim {\cal C}{\cal N}\left( {0,{{\bf{I}}_N}} \right)$.
Based on {\bf Lemma 1}, we have
\begin{equation}
	\begin{split}
		{\rm{tr}}\left( {{ {\bf{B}} }_m}\left[ l \right] \right) + \sqrt {2{\sigma}_2  }\sqrt { \left( {\left\| {{ {\bf{B}} }_m}\left[ l \right] \right\|_F^2 + 2{{\left\| {{ {\bf{B}} }_m}\left[ l \right]{{\widehat {\bf{g}}}_m} \right\|}^2}} \right)} \\
		+ {\sigma}_2 \lambda^+\left({\bf B}_m\left[l\right]\right)
		\le \sigma _{{g,m}}^2,\forall m,\label{t2}
	\end{split}
\end{equation}
Finally, we transform Eq. (\ref{t2}) into
\begin{equation}
	{\rm{tr}}\left( {{ {\bf{B}} }_m}\left[ l \right] \right) + \sqrt {2{\sigma}_2  }\rho_m\left[l\right]  + {\sigma}_2 \mu_m\left[l\right]
	\le \sigma _{{g,m}}^2,\forall m,\label{qutb}
\end{equation}
\begin{equation}
	\left\|\left[ { vec\left({{ {\bf{B}} }_m}\left[ l \right]\right)  ; \sqrt{2}{{ {{ {\bf{B}} }_m}\left[ l \right]{{\widehat {\bf{g}}}_m} }}} \right]\right\|\le \rho_m\left[l\right],\forall m,
\end{equation}
\begin{equation}
	\mu_m\left[l\right]{\bf I}_N-{{ {\bf{B}} }_m}\left[ l \right]\succeq 0,\forall m,\label{elb}		
\end{equation}
where ${\sigma}_2=-\ln\left(P_{out,2}\right)$, $\rho_m\left[l\right]$ and $\mu_m\left[l\right]$ denotes the non-negative slack variable.

\subsubsection{Detecting Probability Constraint}As for constraint (\ref{PD}), the specific expression of the cumulative distribution function of the chi-square distribution ${F_{{\cal X}_{\left( 2 \right)}^2}}$ as the division of the integrals of two transcendental functions is extremely complex, but since ${F_{{\cal X}_{\left( 2 \right)}^2}}$ is a monotonically increasing function, we transform (\ref{PD}) into
\begin{equation}
	p\left[l\right]{\varsigma} _m\left[l\right]\ge {I_m},\forall m,\label{prp}
\end{equation}
where $I_m$ denotes the threshold. The uncertainty term in ${\varsigma} _m$ is characterized by an bound on its error, i.e., ${d_m}\left[ l \right]={{\widehat d_m} + \Delta {d_m}\left[ l \right]}, \left|{\Delta {d_m}\left[ l \right]} \right| \le {D_m}\left[ l \right]$. Combined with the modular character of the complex numbers, Eq. (\ref{prp}) can be finally written as
\begin{equation}
	\frac{p\left[l\right]{S_{RCS}} {\lambda_c}^2\left|{\bf a}^H_m\left[l\right]{\bf \widehat a}_m\left[l\right]\right|^2}{ \sigma _r^2N^2{(4\pi)^3\left({\widehat d_m} +{D_m}\left[ l \right]\right)^4}}\ge {I_m},\forall m.\label{prpf}
\end{equation}

Due to the rank-1 constraint (\ref{rank1}), the problem (P0) is still non-convex. The classical way to make the problem convex is to subtract the rank-1 constraint\cite{10018908}.

\subsubsection{Convex Transformation of The Objective Function}Above, we dealt with all the non-convex constraints and next dealt with the objective function. We first transform the ${\mathop {\min }\limits_{\Delta {{\bf{h}}_k}\left[ l \right]} {R_k}\left[ l \right]}$ into 
\begin{equation}
			\setlength{\abovedisplayskip}{3pt}
	\setlength{\belowdisplayskip}{3pt}
	2^{o_k\left[l\right]-C_k\left[l\right]}-1\le \chi_k\left[l\right] ,\forall k,\label{ok}
\end{equation}
and
\begin{equation}
	\chi_k\left[l\right]\le{\mathop {\min }\limits_{\Delta {{\bf{h}}_k}\left[ l \right]} {\gamma _{p,k}}\left[l\right]},\forall k,\label{sinr}
\end{equation}
where $o_k\left[l\right]\ge0$ and $\chi_k\left[l\right]\ge0$ are the slack variables. Then, we present the following lemma for tackling Eq. (\ref{sinr}).

\begin{lemma}
	S-Procedure {\rm \cite{CVX}:} Let functions $f_i\left({\bm \eta }\right)$, $i \in \left\{1,2\right\}$ be defined as
	\[f_i\left(\bm \eta\right)={\bm \eta}^H{\bm U}_i{\bm \eta}+2{\rm Re}\left\{{\bm u}_i^H{\bm \eta}\right\}+u_i,\]
	where ${\bm \eta } \in \mathbb{C}^{N\times 1}$, ${\bm U}_i \in \mathbb{H}^{N\times N}$, ${\bm u}_i\in\mathbb{C}^{N\times 1}$, and $u_i\in\mathbb{R}$. Then the implication $f_1\left({\bm \eta}\right)\le0 \Rightarrow f_2\left({\bm \eta}\right)\le 0$ holds if and only if there exists a $s \ge 0$ such that
	\[s \left[ {\begin{array}{*{20}{c}}
			{{{\bf{U}}_1}}&{{{\bf{u}}_1}}\\
			{{\bf{u}}_1^H}&{{u_1}}
	\end{array}} \right] - \left[ {\begin{array}{*{20}{c}}
			{{{\bf{U}}_2}}&{{{\bf{u}}_2}}\\
			{{\bf{u}}_2^H}&{{u_2}}
	\end{array}} \right] \succeq 0,\]
\end{lemma}
{\noindent provided that there exists a point $\hat {\bm \eta}$ such that $f_i\left(\hat {\bm \eta}\right)<0$.}

Based on {\bf Lemma 2}, we have
\begin{equation}
			\setlength{\abovedisplayskip}{3pt}
	\setlength{\belowdisplayskip}{3pt}
f_1(\Delta {\bf h}_k\left[l\right])={\Delta {\bf h}_k^H\left[l\right]}{\bf I}_N{\Delta {\bf h}_k\left[l\right]}\le{\varepsilon _k}^2\left[ l \right],
\end{equation}
and
\begin{equation}
			\setlength{\abovedisplayskip}{3pt}
	\setlength{\belowdisplayskip}{3pt}
\begin{split}
	&{f_2}\left( {\Delta {{\bf{h}}_k}}\left[ l \right] \right) = \Delta {\bf{h}}_k^H\left[ l \right]{{\bf{U}}_k}\left[ l \right]\Delta {{\bf{h}}_k}\left[ l \right]
	  \\&+ 2{\mathop{\rm Re}\nolimits} \left\{ {\widehat {\bf{h}}_k^H{{\bf{U}}_k}\left[ l \right]\Delta {{\bf{h}}_k}\left[ l \right]} \right\}+ \widehat {\bf{h}}_k^H{{\bf{U}}_k}\left[ l \right]{\widehat {\bf{h}}_k}  \le - \sigma _{{n_k}}^2{\chi _k}\left[ l \right],
\end{split}
\end{equation}
where ${\bf U}_k\left[l\right]={\chi _k}\left[ l \right]\sum\nolimits_{i \ne k}^K{\bf W}_i\left[l\right]-{\bf W}_k\left[l\right]$. Let ${\vartheta _k}\left[ l \right]\ge0$, thus Eq. (\ref{sinr}) is transformed into (\ref{D}). 
 \begin{figure*}[ht] 
	\centering
	\begin{equation}
		{{\bf{D}}_k}\left[ l \right] = \left[ {\begin{array}{*{20}{c}}
				{{\vartheta _k}\left[ l \right]{{\bf{I}}_N} - {{\bf{U}}_k}\left[ l \right]}&{ - {{\bf{U}}_k}\left[ l \right]{{\widehat {\bf{h}}}_k}}\\
				{ - \widehat {\bf{h}}_k^H{{\bf{U}}_k}\left[ l \right]}&{-\widehat {\bf{h}}_k^H{{\bf{U}}_k}\left[ l \right]{{\widehat {\bf{h}}}_k} - \sigma _{{n_k}}^2{\chi _k}\left[ l \right] - {\vartheta_k}\left[ l \right]{\varepsilon _k}^2\left[ l \right]}
		\end{array}} \right] \succeq {\rm{0}},\label{D}
	\end{equation}
	\hrulefill
\end{figure*}
\newcounter{TempEqCnt1}                         
\setcounter{TempEqCnt1}{\value{equation}} 
\setcounter{equation}{72}    
\begin{figure*}[ht] 
	\centering
	\begin{equation}
		{{\bf{F}}_{k,m}}\left[ l \right] = \left[ {\begin{array}{*{20}{c}}
				{{\zeta  _{k,m}}\left[ l \right]{{\bf{I}}_N} - {{\widetilde{\bf W}}_k\left[ l \right]}}&{ - {{\widetilde{\bf W}}_k\left[ l \right]}{{\widehat {\bf{g}}}_m}}\\
				{ - \widehat {\bf{g}}_m^H{{\widetilde{\bf W}}_k\left[ l \right]}}&{-\widehat {\bf{g}}_m^H{{{\widetilde{\bf W}}_k\left[ l \right]}}{{\widehat {\bf{g}}}_m} + \pi_{k,m}\left[ l \right] - {\zeta_{k,m}}\left[ l \right]{\tau _m}^2\left[ l \right]}
		\end{array}} \right] \succeq {\rm{0}},\label{F}
	\end{equation}
	\hrulefill
\end{figure*}
\setcounter{equation}{\value{TempEqCnt1}} 
\newcounter{TempEqCnt2}                         
\setcounter{TempEqCnt2}{\value{equation}} 
\setcounter{equation}{73}    
\begin{figure*}[ht] 
	\centering
	\begin{equation}
		{{\bf{P}}_{k,m}}\left[ l \right] = \left[ {\begin{array}{*{20}{c}}
				{{\beta _{k,m}}\left[ l \right] + {\sigma_{v_m}^2}\left(4\pi\right)^2\left(1+\kappa_v\right)\iota_k\left[ l \right]}&{ 2{\widehat d_m}{\sigma_{v_m}^2}\left(4\pi\right)^2\left(1+\kappa_v\right)\iota_k\left[ l \right]}\\
				{2{\widehat d_m}{\sigma_{v_m}^2}\left(4\pi\right)^2\left(1+\kappa_v\right)\iota_k\left[ l \right]}&{-{\beta _{k,m}}\left[ l \right]{D_m^2}\left[ l \right]+{\widehat d_m}{\sigma_{v_m}^2}\left(4\pi\right)^2\left(1+\kappa_v\right)\iota_k\left[ l \right]-\lambda_c^2\pi _{k,m}\left[l\right]}
		\end{array}} \right] \succeq {\rm{0}}.\label{P}
	\end{equation}
	\hrulefill
\end{figure*}
\setcounter{equation}{\value{TempEqCnt2}}

Secondly, the term ${\mathop {{\rm{max}}}\limits_{m \in M} {\mathop {\max }\limits_{{\Xi _m}\left[ l \right]} {R_{k,m}}\left[ l \right]}}$ is transformed into
\begin{equation}
			\setlength{\abovedisplayskip}{3pt}
	\setlength{\belowdisplayskip}{3pt}
	\omega_k\left[l\right] \ge \log_2(1+{\iota _k}\left[l\right]),\forall k,\label{omi}
\end{equation}
and
\begin{equation}
			\setlength{\abovedisplayskip}{3pt}
	\setlength{\belowdisplayskip}{3pt}
	{\iota _k}\left[l\right]\ge {\mathop {\max }\limits_{{\Xi _m}\left[ l \right]} {\gamma_{k,m}}\left[ l \right]},\forall k,m,\label{max}
\end{equation}
where $\omega_k\left[l\right]$ and ${\iota _k}\left[l\right]$ are slack variables.
For Eq. (\ref{omi}), ultilize the successive convex approximation (SCA) in the $r$-th iteration.
\begin{equation}
			\setlength{\abovedisplayskip}{3pt}
	\setlength{\belowdisplayskip}{3pt}
	\begin{split}
			&\omega_k\left[l\right]\ge \log_2\left(1+\iota_k\left[l\right]\right)\ge \log_2\left(1+\iota_k^{(r)}\left[l\right]\right)\\
			&~~~~~~~~~~+\frac{1}{\ln\left(2\right)\left(1+\iota_k^{(r)}\left[l\right]\right)}\left(\iota_k\left[l\right]-\iota_k^{(r)}\left[l\right]\right).\label{omiga}
	\end{split}
\end{equation} 
Then, Eq. (\ref{max}) is transformed into
\begin{equation}
			\setlength{\abovedisplayskip}{3pt}
	\setlength{\belowdisplayskip}{3pt}
	\begin{split}
		&{\Delta {\bf g}_{m}^H\left[l\right]}\widetilde{\bf W}_k\left[ l \right]{\Delta {\bf g}_{m}\left[l\right]}+2{\rm Re}\left\{{\widehat{\bf g}}_m^H\widetilde{\bf W}_k\left[ l \right]{\Delta {\bf g}_{m}\left[l\right]}\right\}\\
		&~~~~~~~~~~~~~~~~~~~~~~+{\widehat{\bf g}}_m^H\widetilde{\bf W}_k\left[ l \right]{\widehat{\bf g}}_m \le \pi_{k,m}\left[l\right],\forall k,m, \label{pi}
	\end{split}
\end{equation}
and
\begin{equation}
	\pi _{k,m}\left[l\right]\le \frac{{\sigma_{v_m}^2}\left(1+\kappa_v\right)\iota_k\left[l\right]}{\xi_m^2\left[l\right]}, \left| {\Delta {d_m}\left[ l \right]} \right| \le {D_m}\left[ l \right], \forall k,m, \label{dm}
\end{equation}
where $\widetilde{\bf W}_k\left[ l \right]={\bf W}_k\left[ l \right]-\iota_k\left[l\right]{\bf W}_c\left[ l \right]$ and $\pi _{k,m}\left[l\right]$ denotes the slack variable. Similarly, based on {\bf Lemma 2}, Eqs. (\ref{pi}) and (\ref{dm}) can be transformed as (\ref{F}) and (\ref{P}). ${\zeta  _{k,m}}\left[ l \right]\ge0$ and ${\beta _{k,m}}\left[ l \right]\ge0$ are slack variables. 
\newcounter{TempEqCnt3}                         
\setcounter{TempEqCnt3}{\value{equation}} 
\setcounter{equation}{74} 

\subsubsection{Problem Rewrite}Finally, having tackled the objective function and all non-convex constraints, we rewrite the problem (P0) as
\begin{subequations}
	\begin{align}
		&\left( {{\textrm{P1}}} \right){\textrm{:~}}{\mathop \textrm{max}\limits_{\cal B}}~{\frac{1}{T}}\sum\nolimits_{l = 1}^L {t\left[ l \right]}\sum\nolimits_{k = 1}^K{\left(o_k\left[l\right]-\omega_k\left[l\right]\right)}, \notag\\
		&~~~~~~~~~~{\textrm{s}}{\textrm{.t}}{\rm{.}}~~~{\textrm{(\ref{suc})-(\ref{Ck}), (\ref{beam}), (\ref{co}),}}\\
		&~~~~~~~~~~~~~~~~~{\textrm{(\ref{qut})-(\ref{el}), (\ref{qutb})-(\ref{elb}), (\ref{prpf}), (\ref{ok}),}}\\
		&~~~~~~~~~~~~~~~~~{\textrm{(\ref{D}), (\ref{omiga}), (\ref{F}), (\ref{P}).}}
	\end{align}	
\end{subequations}
where ${\cal B}=\left\{t\left[ l \right], {C_k}\left[ l \right], {{\bf{W}}_k}\left[ l \right], {{\bf{W}}_c}\left[ l \right], {z_k}\left[ l \right], {\nu _k}\left[ l \right], {\rho _m}\left[ l \right], \right.\\$ $\left.{\mu _m}\left[ l \right], {\vartheta _k}\left[ l \right], {\zeta _{k,m}}\left[ l \right], {\beta _{k,m}}\left[ l \right], {\pi _{k,m}}\left[ l \right], {\omega _k}\left[ l \right], {o_k}\left[ l \right], {\chi _k}\left[ l \right], \right.\\$ $\left.{\iota _k}\left[ l \right]\right\}$. Notice that problem (P1) is still a nonconvex problem due to the coupling of the variables, and in the next subsection, it is separated into subproblems for solving.

\subsection{Problem Solving}
In this subsection, due to the coupling of variables, we split the problem (P1) into two subproblems to solve. The BCD-based method is able to efficiently solve high-quality suboptimal solutions to the problem with appropriate computational complexity by alternating iterations. Therefore we divide the optimization variables into 2 blocks and apply the BCD algorithm to solve the problem (P1). 

\subsubsection{Block 1}For the first block ${\cal B}_1=\left\{t\left[l\right],{z_k}\left[ l \right],{\nu _k}\left[ l \right],{\rho _m}\left[ l \right],{\mu _m}\left[ l \right],{\pi _{k,m}}\left[ l \right],{{\bf{W}}_k}\left[ l \right],{{\bf{W}}_c}\left[ l \right]\right\}$, it can be obtained by solving the problem (P2.1) for given the block ${\cal B}_2=\left\{{C_k}\left[ l \right],{\chi _k}\left[ l \right],{\iota _k}\left[ l \right],{o_k}\left[ l \right],{\omega _k}\left[ l \right],{\pi _{k,m}}\left[ l \right],a_k\left[l\right]\right\}$\footnote{Since ${\pi _{k,m}}\left[ l \right]$ is not coupled to any variable, we consider it in two blocks to ensure its joint optimality. The definition of $a_k\left[l\right]$ can be found in the following text.}. Then, the first subproblem can be written as

\begin{subequations}
	\begin{align}
		&\left( {{\textrm{P2.1}}} \right){\rm{:~}}\mathop {{\textrm{max}}}\limits_{{\cal B}_1} {\rm{ ~}}{{\frac{1}{T}}\sum\nolimits_{l = 1}^L {t\left[ l \right]}\sum\nolimits_{k = 1}^K{\left(o_k\left[l\right]-\omega_k\left[l\right]\right)}, } \notag\\
		&~~~~~~~~~~~~~{\textrm{s}}{\textrm{.t}}{\rm{.}}~~~{\textrm{(\ref{suc})-(\ref{C0}), (\ref{beam})}},\\
		&~~~~~~~~~~~~~~~~~~~~{\textrm{(\ref{qut})-(\ref{el}), (\ref{qutb})-(\ref{elb}), (\ref{prpf})}},\\
		&~~~~~~~~~~~~~~~~~~~~{\textrm{(\ref{D}), (\ref{F}), (\ref{P})}}.
	\end{align}	
\end{subequations}
The constraints of problem (P1.1) contain linear matrix inequalities (LMI) and second-order cones (SOC), which can be well solved by the SOCP algorithm\cite{9234527,6891348}.

\subsubsection{Block 2}Since constraints (\ref{qut}), (\ref{soc}), and (\ref{el}) are nonconvex with respect to the variable ${C_k}\left[ l \right]$, we can not obtain the second block ${\cal B}_2=\left\{{C_k}\left[ l \right],{\omega _k}\left[ l \right],{\chi _k}\left[ l \right],{\iota _k}\left[ l \right],{o_k}\left[ l \right],{\pi _{k,m}}\left[ l \right],a_k\left[l\right]\right\}$. We transform the constraints into the following form
\begin{equation}
	{C_k}\left[ l \right] \ge r_k^s-\log_2(\frac{{\sigma_{h_e}^2}{{\widehat {\bf{h}}}_k^H}{\bf{W}}_k\left[ l \right]{{\widehat {\bf{h}}}_k}}{{\sigma_{n_k}^2}-tp_{1,k}\left[l\right]+tp_{2,k}\left[l\right]}+1),\forall k,\label{tp}
\end{equation}
\begin{equation}
	\left\|\left[ { vec\left({{\widetilde {\bf{A}} }_k}\left[ l \right]\right)  ; \sqrt{2}{{ {{\widetilde {\bf{A}} }_k}\left[ l \right]{{\widehat {\bf{h}}}_k} }}} \right]\right\|  \le z_k\left[l\right],\forall k,\label{norm}
\end{equation}
\begin{equation}
	a_k\left[l\right]\ge\frac{1}{2^{r_k^s-C_k\left[l\right]}-1},\forall k,\label{tou}
\end{equation}
\begin{equation}
		2^{{C_k}\left[ l \right]-r_k^s}+{\bf I}_N+{\bf M}^{-1}\left[l\right]\succeq 0,\forall k,\label{se}
\end{equation}
For the sake of simplicity, we define the substitutions $tp_{1,k}\left[l\right]={\rm{tr}}\left(\sum\nolimits_{i \ne k}^K{\bf{W}}_i\left[ l \right]\right)-\sqrt {2{\sigma}_1  }z_k\left[l\right]-{\sigma}_1 \nu_k\left[l\right]$, $tp_{2,k}\left[l\right]={\sigma_{h_e}^2}{{\widehat {\bf{h}}}_k^H}\sum\nolimits_{i \ne k}^K{\bf{W}}_i\left[ l \right]{{\widehat {\bf{h}}}_k}$, ${{\widetilde {\bf{A}} }_k}\left[ l \right]={\sigma_{h_e}^2}\left(a_k\left[l\right]{\bf{W}}_k\left[ l \right]-\sum\nolimits_{i \ne k}^K{\bf{W}}_i\left[ l \right]\right)$, $a_k\left[l\right]$ denotes the auxiliary variable, and ${\bf M}\left[l\right]=\left(\sum\nolimits_{i \ne k}^K{\bf{W}}_i\left[ l \right]-\nu_k\left[l\right]{\bf I}_N\right){\bf{W}}_k^{-1}\left[ l \right]$. Noting that Eq. (\ref{tou}) is not easy to handle, we transform it into 
\begin{equation}
			\setlength{\abovedisplayskip}{3pt}
	\setlength{\belowdisplayskip}{3pt}
	r_k^s-C_k\left[l\right]\ge\log_2\left(\frac{1}{a_k\left[l\right]}+1\right),\forall k.\label{per}
\end{equation}
Evidently, it is a perspective function. Therefore, the block ${\cal B}_2=\left\{{C_k}\left[ l \right],{\omega _k}\left[ l \right],{\chi _k}\left[ l \right],{\iota _k}\left[ l \right],{o_k}\left[ l \right],{\pi _{k,m}}\left[ l \right],a_k\left[l\right]\right\}$ can be obtained by solving the problem (P2.2) for given ${\cal B}_1=\left\{t\left[l\right],{z_k}\left[ l \right],{\nu _k}\left[ l \right],{\rho _m}\left[ l \right],{\mu _m}\left[ l \right],{\pi _{k,m}}\left[ l \right],{{\bf{W}}_k}\left[ l \right],{{\bf{W}}_c}\left[ l \right]\right\}$. Then, the second optimization problem is given as follow

\begin{subequations}
	\begin{align}
	&\left( {{\textrm{P2.2}}} \right){\rm{:~}}\mathop {{\textrm{max}}}\limits_{{\cal B}_2} {\rm{ ~}}{{\frac{1}{T}}\sum\nolimits_{l = 1}^L {t\left[ l \right]}\sum\nolimits_{k = 1}^K{\left(o_k\left[l\right]-\omega_k\left[l\right]\right)}, } \notag\\
	&~~~~~~~~~~~~~{\textrm{s}}{\textrm{.t}}{\rm{.}}~~~{\textrm{(\ref{Ck}), (\ref{C0}), (\ref{ok}), (\ref{D}), (\ref{omiga})}},\\
	&~~~~~~~~~~~~~~~~~~~~{\textrm{(\ref{F}), (\ref{P}), (\ref{tp}), (\ref{norm}), (\ref{se}), (\ref{per})}}.
\end{align}	
\end{subequations}
The constraints of problem (P2.2) contain linear inequalities and LMI, so it can be solved by the interior point method\cite{cvxtool}. 

As above mentioned, (P2.1) and (P2.2) are alternately iterated to find the optimal solution to problem (P1), which can be summarized as the robust secure ISAC design algorithm shown in {\bf Algorithm \ref{alg1}}.

\begin{algorithm}[htbp]
	\caption{Robust and Secure ISAC Design Algorithm}
	\label{alg1}
	\begin{algorithmic}[1]
		\STATE {\bf{Initialization}}: ${\cal B}_1^{\left(0\right)}$, ${\cal B}_2^{\left(0\right)}$, $\forall l \in {\cal L}$, convergence threshold $\varepsilon$ and iteration index $r = 0$.
		\REPEAT
		\STATE Obtain block ${\cal B}_1^{\left(r\right)}$ by solving problem (P2.1) utilizing the SOCP.

		\STATE Obtain block ${\cal B}_2^{\left(r\right)}$ by solving problem (P2.2) utilizing the interior point method.
		\STATE $r \leftarrow r + 1$.
		\UNTIL The fractional decrease of the objective value is
		below a threshold $\varepsilon$.
		\STATE {\bf{return}} Beamforming matrix ${\bf{W}}\left[l\right]$, common stream vector ${\bf c}\left[l\right]$, and timeslot duration $t\left[l\right]$.
	\end{algorithmic}  
\end{algorithm}

\subsection{Convergence and Computational Complexity Analysis}
{\it 1) Computational Complexity Analysis:} The problem (P2.1) is solved with complexity ${\cal O}\left( {KLMN^{4.5}} \right)$, and the problem (P2.2) is solved with complexity ${\cal O}\left( {KLM}N^{3.5} \right)$. Therefore, the overall computational complexity of {\bf Algorithm \ref{alg1}} is ${\cal O}\left( {KLM\log \left( {1/\varepsilon } \right)N^{4.5}} \right)$\cite{0Lectures}, where $\varepsilon$ denotes the precision of stopping the iteration.

{\it 2) Convergence Analysis:} The convergence analysis of {\bf Algorithm \ref{alg1}} can be proved as follows. We define ${\cal B}_1^{\left(r\right)}$, and ${\cal B}_2^{\left(r\right)}$ as the solutions of the $r$-th iteration of problems (P2.1), and (P2.2), so the objective function of the $r$-th iteration can be expressed as ${\cal F}\left( {\cal B}_1^{\left(r\right)},{\cal B}_2^{\left(r\right)} \right)$.
In step 3 of {\bf Algorithm \ref{alg1}}, the block ${\cal B}_1^{\left(r+1\right)}$ can be obtained given the block ${\cal B}_2^{\left(r\right)}$. Then there are
\begin{equation}
		\begin{split}
		{\cal F}\left( {\cal B}_1^{\left(r\right)},{\cal B}_2^{\left(r\right)} \right)\le {\cal F}\left( {\cal B}_1^{\left(r+1\right)},{\cal B}_2^{\left(r\right)} \right).
		\end{split}
\end{equation}
In step 4 of {\bf Algorithm \ref{alg1}}, the ${\cal B}_2^{\left(r+1\right)}$ can be obtained with ${\cal B}_1^{\left(r+1\right)}$ given, then there are

\begin{equation}
	\begin{split}
	{\cal F}\left({\cal B}_1^{\left(r+1\right)},{\cal B}_2^{\left(r\right)}  \right)\le {\cal F}\left( {\cal B}_1^{\left(r+1\right)},{\cal B}_2^{\left(r+1\right)} \right).
	\end{split}
\end{equation}
Based on the above, there is ultimately
\begin{equation}
	\begin{split}
	{\cal F}\left({\cal B}_1^{\left(r\right)},{\cal B}_2^{\left(r\right)}  \right)\le {\cal F}\left( {\cal B}_1^{\left(r+1\right)},{\cal B}_2^{\left(r+1\right)} \right).
 	\end{split}
\end{equation}
This shows that at each iteration of {\bf Algorithm \ref{alg1}}, the objective function is non-decreasing. Since the objective function must be finite-valued upper bound, the convergence of {\bf Algorithm \ref{alg1}} can be guaranteed. Since the upper and lower bounds of the constraints or functions are obtained in the algorithm by using a relaxation process such as SCA, the solution obtained by solving is a suboptimal solution with high accuracy.
\section{Numerical Results}
In this section, numerical simulations of the proposed algorithm are performed to demonstrate its effectiveness. A three-dimensional polar coordinate system is used, the BS is located at $\left(0\rm{m}, 0^{\circ}, 0^{\circ}\right)$, $K=3$ LUs are distributed at coordinates $\left(50\rm{m}, 22.5^{\circ}, 10^{\circ}\right)$, $\left(70\rm{m}, 45^{\circ}, 20^{\circ}\right)$, and $\left(90\rm{m}, 67.5^{\circ}, 30^{\circ}\right)$, respectively, and $M=2$ IUs are distributed near the LUs, with coordinates of $\left(65\rm{m}, 35^{\circ}, 25^{\circ}\right)$ and $\left(85\rm{m}, 55^{\circ}, 35^{\circ}\right)$, respectively. The remaining parameters are given in Table I.
\begin{table}[!htbp]
	\caption{Simulation Parameters.}
	\begin{center}
\begin{tabular}{|l|c|}  
	\hline
	\bf{Parameters} & \bf{Value}\\ 
	\hline
	Carrier frequency ($f$) & 30 GHz \\
	\hline
	Maximum transmissive power ($P_{t}$) & 1 mW \\
	\hline
	System bandwidth ($W$) & 20 MHz \\
	\hline
	Minimum duration (${t_{min}}$) & 0.5 ms \\
	\hline
	Maximum duration (${t_{max}}$) & 3.5 ms \\
	\hline
	Scanning period (${T}$) & 10 ms \\
	\hline
	Noise power (${\sigma_{n_k} ^2}$, ${\sigma_{v_m} ^2}$) & -90 dBm \\
	\hline
	Convergence precision ($\varepsilon$) & $10^{ - 3}$\\
	\hline
	LU outage probability rate threshold ($r_k^s$) & $0.5$ bps/Hz\\
	\hline
	IU outage probability rate threshold ($r_m^e$) & $0.25$ bps/Hz\\
	\hline
	LU outage probability ($P_{out,1}$) & $2\%$\\
	\hline
	IU outage probability ($P_{out,2}$) & $1\%$\\
	\hline
\end{tabular}
	\end{center}
\end{table}

In this paper, we compare the performance of the proposed algorithm and othe benchmarks as follows: (1) {\bf{Traditional Transceiver with RSMA (TRSMA)}}: this scheme uses a traditional multi-antenna transceiver whose power constraint can be expressed as $\sum\nolimits_{i = 1}^{\cal \hat K}{\rm tr}\left({\bf W}_i\left[l\right]\right)\le NP_{t}$. (2) {\bf{Zero Forcing (ZF)}}: this scheme solves the problem (P0) by jointly designing the timeslot duration, the covariance matrix of AN, and the transmit power by utilizing the ZF beamforming and replacing the common stream with AN. (3) {\bf{Semidefinite Programming (SDP)}}: this scheme utilizes the SDP algorithm and replaces the common stream with AN and the outage constraints with secrecy sum-rate threshold constraints. (4) {\bf{Space Division Multiple Access (SDMA)}}: this scheme utilizes SDMA as the access method and introduces AN. (5) {\bf{Non-Orthogonal Multiple Access (NOMA)}}: this scheme utilizes NOMA as the access method and introduces AN.
\begin{figure}[H]
	\centerline{\includegraphics[width=6.0cm]{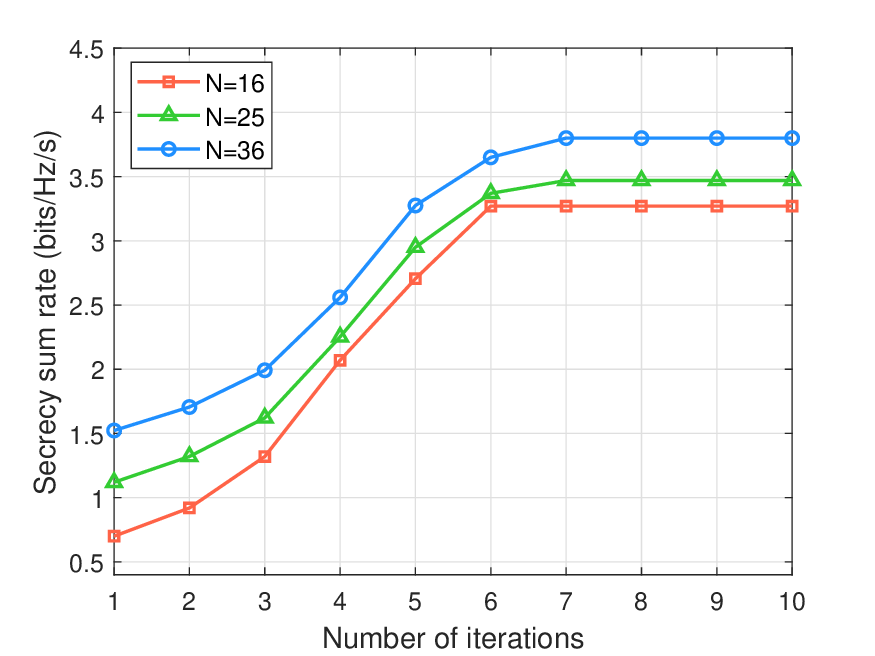}}
	\caption{Convergence process: secrecy sum-rate versus number of iterations under different TRIS elements (${\varepsilon _k}=\sqrt{0.01}, {\varepsilon _{m,n}}=0.1\sqrt{\kappa_v}, P_t = 1{\rm mW}$).}\label{conv}
\end{figure}
First, we verify the convergence of the proposed robust and secure ISAC design algorithm. Fig. \ref{conv} depicts the variation of the objective function value with the number of iterations for different TRIS elements. It is clear that the algorithm can achieve a good convergence performance in about 7 iterations. Moreover, the higher the number of TRIS elements, the higher the value of the objective function. This indicates that the secrecy sum-rate of the system increases with the number of TRIS elements.
\begin{figure}[H]
	\centerline{\includegraphics[width=6.0cm]{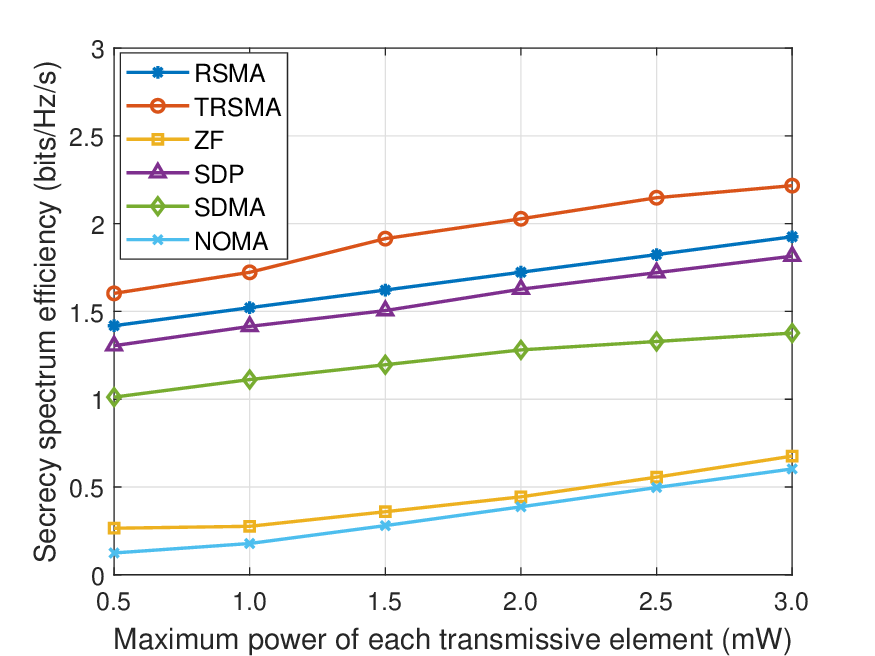}}
	\caption{SSE varies with the maximum power of each transmissive element (${\varepsilon _k}=\sqrt{0.01}, {\varepsilon _{m,n}}=0.1\sqrt{\kappa_v},N=16$).}\label{ssept}
\end{figure}
Secondly, we illustrate the secrecy spectral efficiency (SSE) varies with the maximum power of each transmissive element. As shown in Fig. \ref{ssept}, the SSE of the system increases as the maximum power of each TRIS element increases. Moreover, the SSE of the proposed architecture is second only to the traditional transceiver due to the fact that TMA constrains the rows of the ${\bf{W}}\left[ l \right]$ and actually consumes less energy. Besides, the traditional transceiver has more RF links, which can be independently manipulated with more freedom, and has better accuracy in beamforming, so the SEE is slightly higher than that of the TRIS transceiver.

\begin{figure}
	\centering
	\begin{minipage}[t]{1\linewidth} 
		\centering
		\begin{tabular}{@{\extracolsep{\fill}}c@{}c@{}@{\extracolsep{\fill}}}
			\includegraphics[width=0.54\linewidth]{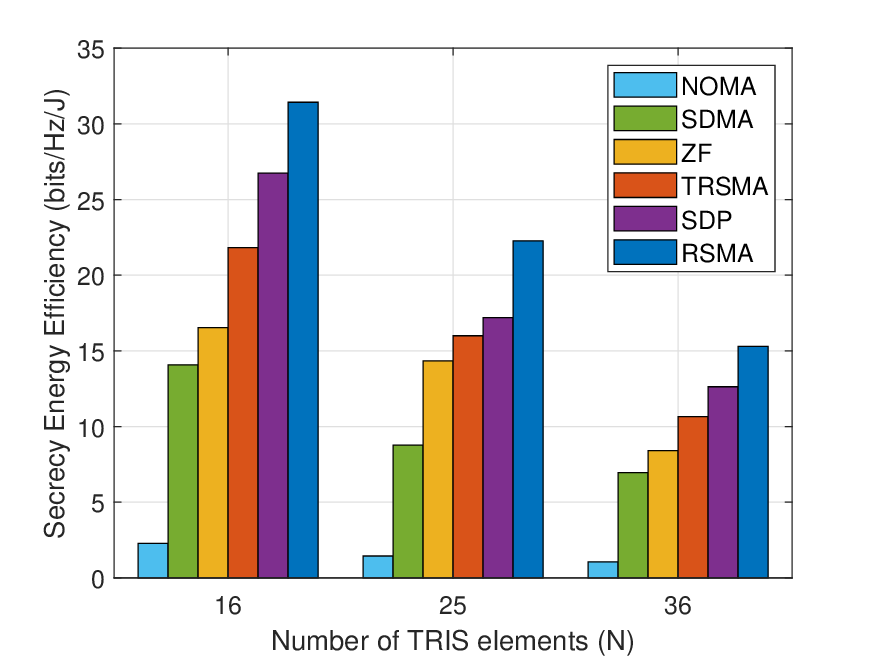}&
			\includegraphics[width=0.54\linewidth]{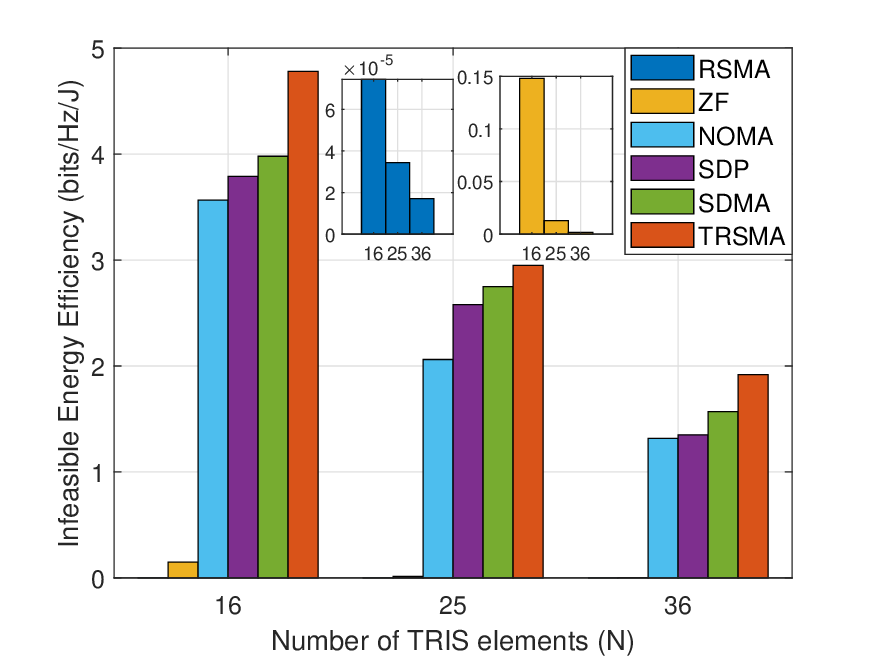}\\
			(a) SEE & (b) IEE\\
		\end{tabular}
	\end{minipage}
	\caption{SEE and IEE varies with the number of TRIS element (${\varepsilon _k}=\sqrt{0.01}, {\varepsilon _{m,n}}=0.1\sqrt{\kappa_v}, P_{t}=1{\rm mW}$).}\label{ab}
\end{figure}
\begin{figure}[!htbp]
	\centerline{\includegraphics[width=6.0cm]{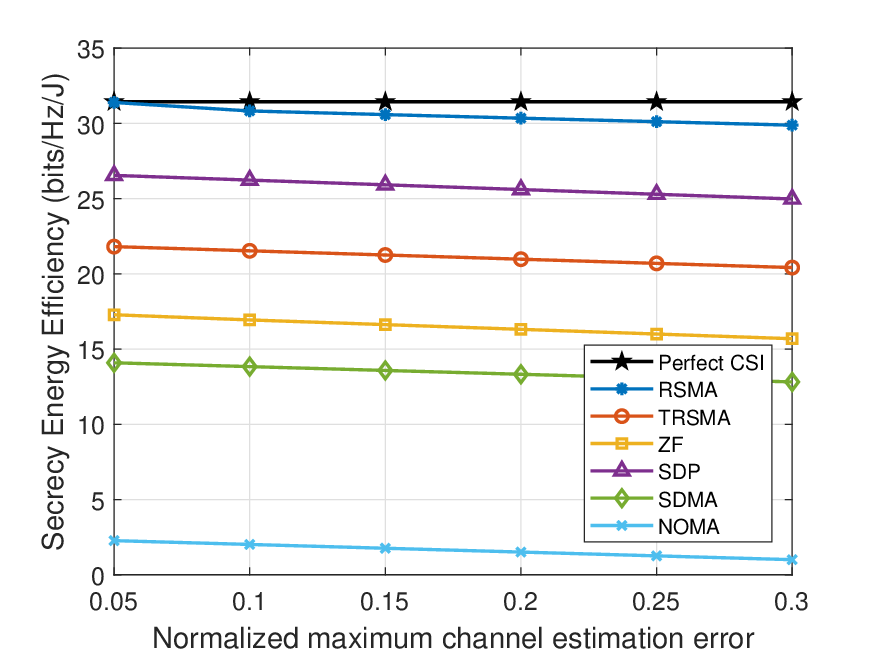}}
	\caption{SEE varies with the normalized channel estimation error $\tau_m^2$ (${\varepsilon _k}=\sqrt{0.01},P_t=1{\rm mW},N=16$).}\label{seeer}
\end{figure}
\begin{figure}[!htbp]
	\centerline{\includegraphics[width=6.0cm]{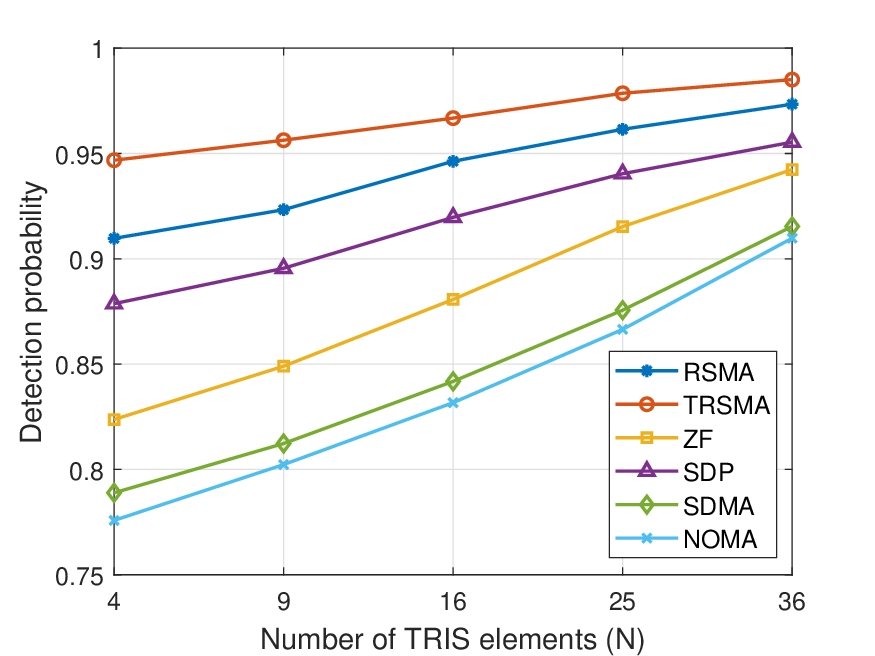}}
	\caption{The detection probability of IU varies with the number of TRIS elements (${\varepsilon _{m,n}}=0.1\sqrt{\kappa_v}, P_t=1{\rm mW}$).}\label{dpn}
\end{figure}
However, the SEE of the proposed scheme outperforms all the benchmarks due to its flexible interference management and simple decoding sequence as shown in Fig. \ref{ab} (a), which confirms the superiority of the TRIS transceiver in terms of SEE, as well as the 44\% improvement in SEE compared to the traditional transceiver. The SE of the proposed scheme is comparable to that of the traditional transceiver, but its actual energy consumption is lower compared to the traditional scheme due to the row constraints on the precoded matrices, and thus the SEE is higher than the baseline. And according to Fig. \ref{ab} (b), the infeasible energy efficiency (IEE) of the proposed scheme outperforms all the benchmark schemes and decreases with the increase in the number of TRIS elements, confirming the effectiveness of increasing the number of TRIS elements in reducing eavesdropping by IUs, where IEE is the ratio of the eavesdropped sum rate to the total power. In addition, SEE decreases as the number of TRIS elements increases, which is due to the increase in power consumption with the increase in TRIS elements, but the enhancement for the secrecy rate is minor.
\begin{figure*} [!htbp]
	\centering
	\begin{minipage}[t]{1.0\linewidth}
		\centering
		\begin{tabular}{@{\extracolsep{\fill}}c@{}c@{}c@{}@{\extracolsep{\fill}}}
			\includegraphics[width=0.32\linewidth]{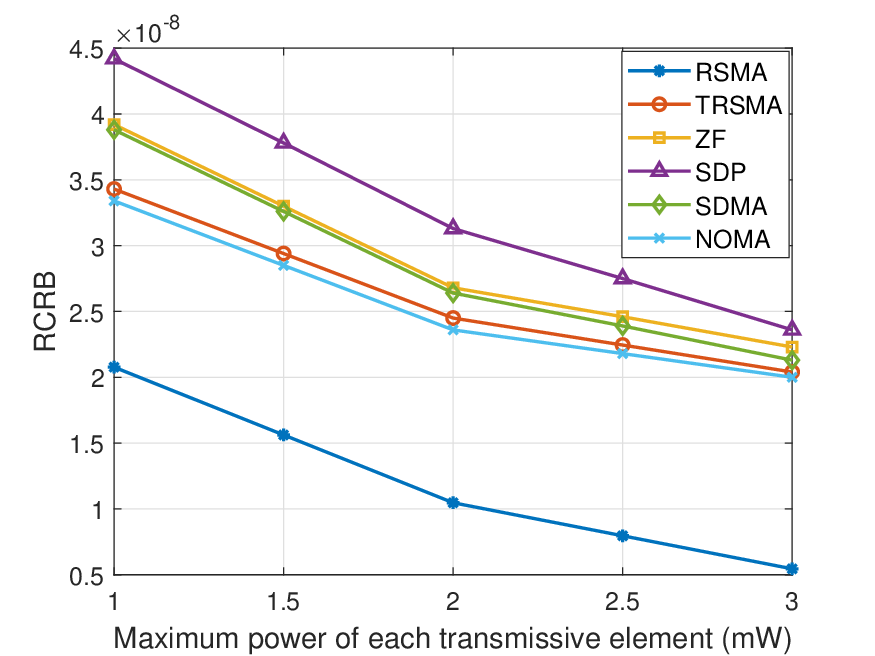}&
			\includegraphics[width=0.32\linewidth]{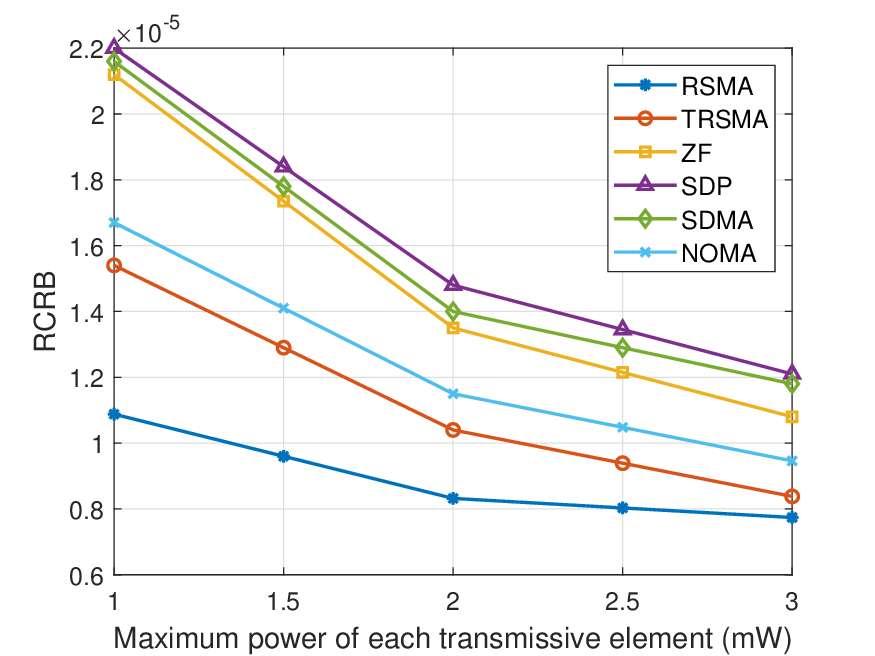}&
			\includegraphics[width=0.32\linewidth]{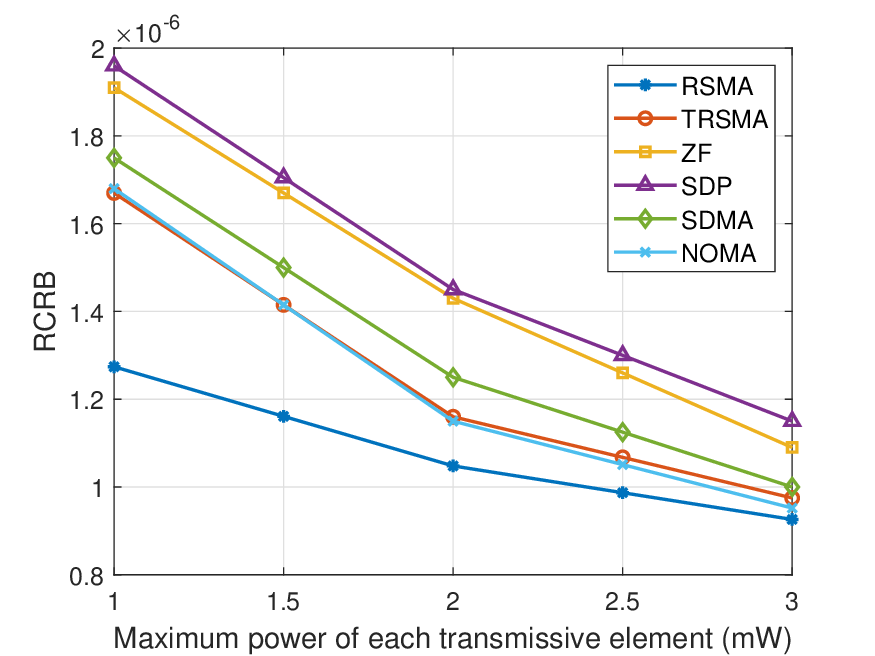}\\
			(a) $d$ RCRB & (b) $\theta$ RCRB & (c) $\phi$ RCRB\\
		\end{tabular}
	\end{minipage}
	\caption{RCRB varies with the maximum power of each transmissive element ($P_{t}=1{\rm mW}, N=16$).}
	\label{crb}
\end{figure*}
\begin{figure*} [!htbp]
	\centering
	\begin{minipage}[t]{1.0\linewidth}
		\centering
		\begin{tabular}{@{\extracolsep{\fill}}c@{}c@{}c@{}@{\extracolsep{\fill}}}
			\includegraphics[width=0.25\linewidth]{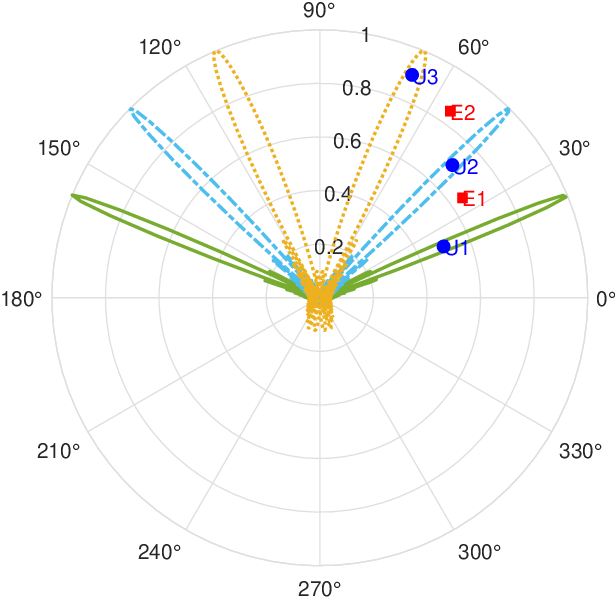}&
			\includegraphics[width=0.25\linewidth]{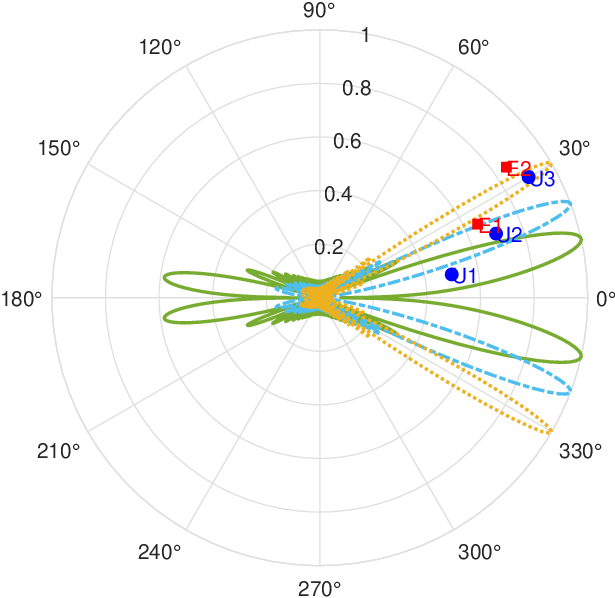}&
			\includegraphics[width=0.33\linewidth]{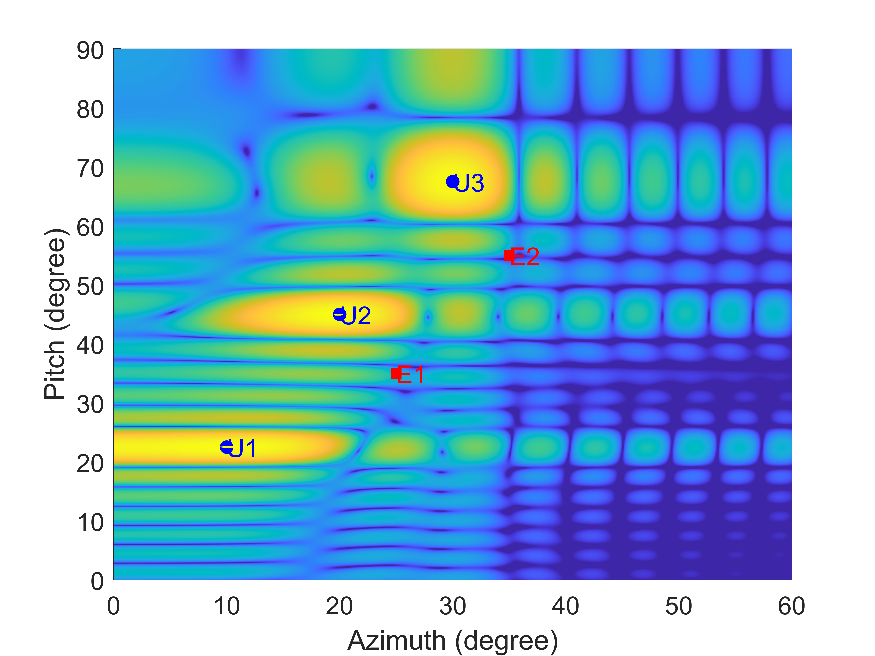}\\
			(a) Pitch & (b) Azimuth  & (c) Beampattern \\
		\end{tabular}
	\end{minipage}
	\caption{Beam scanning and beampattern ($P_{t}=1{\rm mW}, N=36$).}
	\label{Beampattern}
\end{figure*}

Subsequently, in order to explore the effect of channel estimation error on the system's SEE, it is demonstrated in Fig. \ref{seeer}. The perfect CSI baseline is the same design as the RSMA scheme except that the channel utilizes perfect CSI. It can be seen that compared to perfect CSI condition under the proposed algorithm, the SEE of the proposed robust scheme decreases as the channel error increases, has a 5\% performance loss, but outperforms other schemes.

Next, we compare the changes in the detection probability of IU with the increase in the number of TRIS elements for each scheme as shown in Fig. \ref{dpn}. The IU detection probability of the proposed scheme is above 90\% and higher than that of the baseline schemes except for the traditional transceivers, which is due to the fact that the traditional transceiver utilizes multiple receive antennas, so the echo signals it obtain are spatially richer than the TRIS transceiver with only a single receive antenna, and the channel gain for the multiple receive antennas is greater, so the SNR of its echo signals is higher, resulting in a higher probability of detection. To address this drawback, we can deploy multiple low-power sensors at the TRIS receiver to enhance the SNR of the echo signal, which is our further work. Moreover, the detection probability increases with the number of TRIS elements, which is due to the fact that more TRIS elements provide better capture of the echo signals and improve the SNR of the echo signals, which in turn improves the detection probability, which provides guidance for the design of a secure ISAC network.
\begin{figure}[H]
	\centering
	\begin{minipage}[!htbp]{1.0\linewidth}
		\centering
		\begin{tabular}{@{\extracolsep{\fill}}c@{}c@{}@{\extracolsep{\fill}}}
			\includegraphics[width=0.52\linewidth]{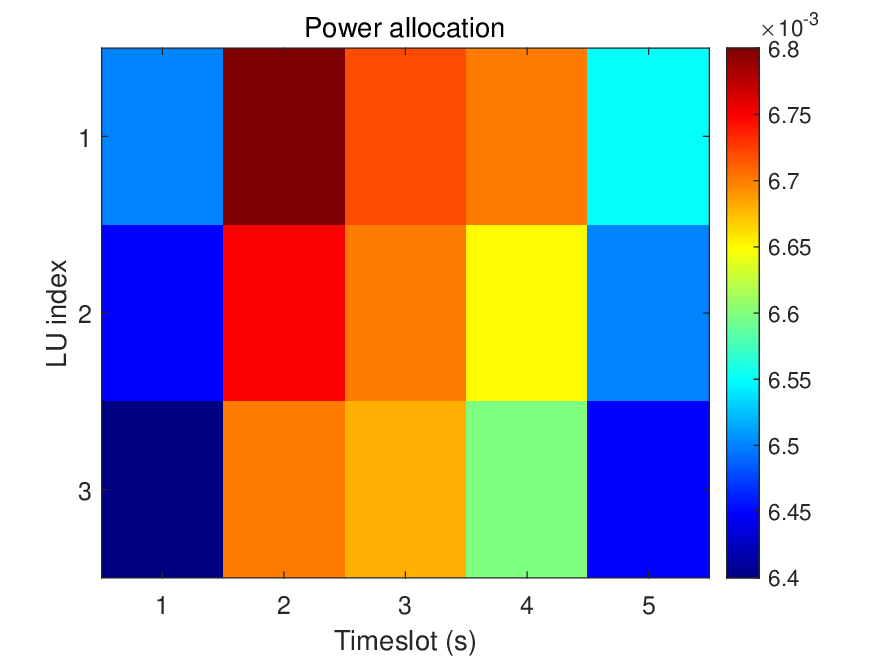}&
			\includegraphics[width=0.52\linewidth]{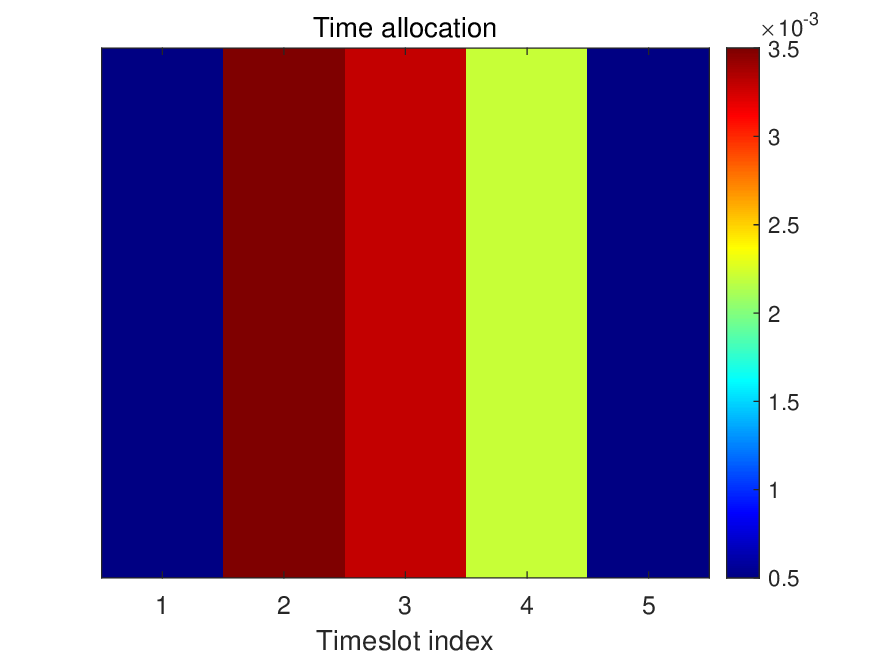}\\
			(a) Power allocation & (b) Timeslot duration allocation\\
		\end{tabular}
	\end{minipage}
	\caption{Resource allocation ($P_{t}=1{\rm mW}, N=36$).}\label{ta}
\end{figure}

Then, in order to compare the impact of each scheme on the sensing accuracy of IUs, we use root Cram\'{e}r-Rao boundary (RCRB) as a criterion\cite{10522473}. From Fig. \ref{crb}, the results show that RSMA outperforms the other schemes in terms of RCRB for distance, azimuth, and pitch, indicating that RSMA has a better ability to manage interference and adapt to radar sensing. 

Next, in order to visualize the beampatterns, we show in Fig. \ref{Beampattern} the beam scanning for timeslots 2, 3, and 4 as well as the beampattern. It is obvious that the beam of the proposed architecture in this paper can well cover the LUs, who are in the peak position of the beampattern, and the IUs are in a region of low beam energy.

Finally, in order to visualize the resource allocation throughout the service process, it is presented in Fig. \ref{ta}. It can be seen that most of the system's power is allocated to timeslots 2, 3, and 4, and that the timeslots have the longest duration, matching the beam scanning results. Beam scanning over time slots is always beneficial, and short beam durations at IUs are effective in avoiding eavesdropping.

\section{Conclusions}
In this paper, we propose a robust and secure ISAC architecture based on time-division. To facilitate the integration of sensing and communication, we deploy a novel TRIS transceiver framework and innovatively exploits the common stream of RSMA as both useful signals and AN to address the problem of IU eavesdropping and interference. Based on the architectural setup, we consider the problem of secure communication and potential IUs detection under conditions of imperfect CSI and networks serving multiple LUs with the presence of multiple IUs and give theoretical upper bounds on the error of IU channels. Also, to improve the security of the system, we consider the system outage. Based on these settings, we propose a joint robust and secure communication and sensing design algorithm. Numerical simulations validate the effectiveness of the proposed architecture and its advantages over other schemes, not only due to the superiority of the proposed transceiver but also the gain of the designed algorithm. In addition, design guidelines are provided for future secure ISAC networks, i.e., increasing the number of TRIS elements, adopting RSMA, and adopting multi-timeslot beam scanning will result in better sensing and secure communication performance enhancement.

\bibliographystyle{IEEEtran}
\bibliography{IEEEabrv,reference}

%
%
%
%
%
%
%
%

\end{document}